\RequirePackage[2020-02-02]{latexrelease}
\RequirePackage{filecontents}

\documentclass{article}
\usepackage{amssymb}
\usepackage{amsbsy}
\usepackage{amsfonts}
\usepackage{amsmath}
\usepackage[margin=1.0in]{geometry}
\usepackage{float}
\usepackage{color}
\usepackage{bm}
\usepackage{csvsimple}
\usepackage{booktabs} 
\usepackage{caption} 
\usepackage{subcaption} 
\usepackage{graphicx}
\usepackage{pgfplots}
\usepackage[all]{nowidow}
\usepackage[utf8]{inputenc}
\usepackage[breaklinks]{hyperref}
\hypersetup{
  colorlinks = true,
}
\newcommand{\mean}[1]{\langle#1\rangle}  
\newcommand{\TC}{\mathcal{T}}  
\newcommand{\DTC}{\mathcal{D}}  

\begin{document}
\title{Pairwise and high-order dependencies in the cryptocurrency trading network}

\author{T. Scagliarini$^{1}$\footnote{Corresponding author: tomas.scagliarini@uniba.it} , G. Pappalardo$^{2}$, A.E. Biondo$^{3}$, \\A. Pluchino$^{2,}$$^{4}$, A. Rapisarda$^{2,}$$^{4,}$$^{5}$, S. Stramaglia$^{1,}$$^{6}$\\
}
\date{%
$^{1}$ Dipartimento Interateneo di Fisica, Università degli Studi Aldo Moro, Bari and INFN, Italy\\
$^{2}$ Dipartimento di Fisica e Astronomia, Università degli Studi di Catania, Italy\\
$^{3}$ Dipartimento di Economia e Impresa, Università degli Studi di Catania, Italy\\
$^{4}$ INFN Sezione di Catania, Italy\\
$^{5}$ Complexity Science Hub, Vienna\\
$^{6}$ Center of Innovative Technologies for Signal Detection and Processing (TIRES), \\Università degli Studi Aldo Moro, Italy
}

\maketitle

\begin{abstract}

In this paper we analyse the effects of information flows in cryptocurrency markets.
We first define a cryptocurrency trading network, i.e. the network made using cryptocurrencies as nodes and the Granger causality among their weekly log returns as links, later we analyse its evolution over time.
In particular, with reference to years 2020 and 2021, we study the logarithmic US dollar price returns of the cryptocurrency trading network using both pairwise and high-order statistical dependencies, quantified by Granger causality and O-information, respectively. 
With reference to the former, we find that it shows peaks in correspondence of important events, like e.g., Covid-19 pandemic  turbulence or occasional sudden prices rise. The corresponding network structure is rather stable, across weekly time windows in the period considered and the coins are the most influential nodes in the network. In the pairwise description of the network, stable coins seem to play a marginal role whereas, turning high-order dependencies, they appear in the highest number of synergistic information circuits, thus proving that they play a major role for high order effects.
With reference to redundancy and synergy with the time evolution of the total transactions in US dollars, we find that their large volume in the first semester of 2021 seems to have triggered a transition in the cryptocurrency network toward a more complex dynamical landscape.  
Our results show that pairwise and high-order  descriptions of complex financial systems provide complementary information for cryptocurrency analysis.

\end{abstract}

\section{Introduction} \label{Introduction}

The interest in cryptocurrencies increased a lot in the last few years, among investors and researchers. Still recently, until approximately five-seven year ago, the cryptocurrency markets were the object of interest of few professionals, vastly academic, interested on applications, technology and even speculation on DeFi (Decentralized Finance). As data shows, during the year 2017, a huge increase in the USD price of Bitcoin (something like $2300 \%$, from about 800 in January to more than 19000 in December \cite{yahoofinance1}) caused un undoubtedly understandable interest.  An even more impressive piece of evidence was that, after just after one month, that same price fell down to $6852$ USD - a fall of more than $64 \%$ \cite{yahoofinance2}. Also other cryptocurrencies manifested what we could name ``turbulence'' (at best) but the very peculiar aspect of their dynamics was the impressive volatility, the apparently intrinsic instability of these financial assets. 

Needless to say, their markets have become attractive as a very ambitious and dangerous lottery, still able to provide wide returns, at a scale difficultly comparable with ``traditional'' financial instruments.  Since then, a proliferous set of cryptocurrencies has developed and traded on exchanges (ETH since 2015, REP, XRP and ETC since 2016, USDT, XLM and BCH since 2017 among many others, see table \ref{tab:kraken} for a fast overview), thus giving rise to an entire ``sector'', as if they should behave similarly, all apparently representing the hottest opportunity of finance.

In time, therefore, the predominant position of Bitcoin has diminished and the financial potential of these cryptos is now so fragmented over a numerous set of competing alternatives, that it is very unlikely that one of them can assume any truly monetary role. One of the apparent elements of appeal of cryptos is the absence of a centralized control over them. The basic idea of the blockchain rests more on the side of tracking transactions: encryption prevents any personal identification of market participants and this attracts many forms of traders, investors, negotiators in a wide variety of clear and unclear market transactions. Therefore, the market capitalization of the crypto world has assumed notable dimensions going to $\$14$ billion in 2014 to $\$831$ billion in 2018, still remaining very high after severe reductions in 2018, until 2020 at $\$192$ billion. On the aggregate side, it is easy to collect data of cryptocurrencies, since most of the anonymized exchanges are often publicly available. 

Crises show periodically that financial markets are complex systems where interactions among individual participants is much more relevant in terms of dynamics that the individual characteristics of individual participants. One of the most relevant reasons why micro-prudential supervision (e.g., Basel Agreements) is almost vein in order to prevent the systemic risk is that it fails to consider complexity features emerging from interconnectedness among financial institutions and their resulting financial network. Then, given the wild and unpredictable instability of the price of cryptocurrencies, a branch of literature has flourished dealing with analysis of signals and precursors that could be useful to check the existence of correlations and systemic regularities. 

In this work, we study the impact of cryptocurrency price returns using statistical tools to detect the flows of information between time series. To this purpose, we first show how, in general, two variables exchange information in a pairwise way by Granger causality, which detects the information flow from a source variable to a target one. 

Granger causality is a statistical method aimed to quantify the gain in linear predictability between time series. The idea of defining causality as a useful tool for predictions \footnote{This definition does not necessarily imply a true causal link between time series, due to the possible presence of confounding or unobserved variables \cite{pearl_causality_2009}. For this reason, in this context causality must be intended as a form of predictive causality.} was introduced by Wiener \cite{wiener_modern_1956} and was later implemented by Granger \cite{granger_economic_1963,granger_investigating_1969} in the context of linear autoregressive models, that will be used in the present work as weight on a directed cryptocurrency traded network where each nodes is a cryptocurrency and each link states if a crypto may be useful on the price prediction of another one. Since 1960, Grange Causality has become a standard tool in econometrics and has recently received a lot of attention in many other contexts, such as neuroscience \cite{stokes2017study} and climatology \cite{kodra2011exploring}.
All these studies take into account the information exchange between pairs of variables, possibly conditioning over a set of confounding elements to remove spurious effects. Recently, it has been shown that in many systems a simple pairwise approach may not fully describe the entire multivariate structure of information exchange, since higher-order effects play an important role in their description \cite{mediano2022integrated}. 

High-order effects might be appreciated under two points of view: when considering multiplets of variables, the joint effect can be greater than the sum of single contributions, a condition known as \emph{synergy}; alternatively, the case in which the overall effect results lower than the sum of the parts, is called \emph{redundancy}. Stated differently, redundancy occurs when multiple copies of the same information can be found in different parts of the system, while synergy refers to the part of the information that is not stored in any specific element, but rather in the joint state  of a group of variables. A trivial example is the case where $X_1$ and $X_2$ are independent binary random variables and $X_3 = \texttt{XOR}(X_1,X_2)$. In this situation, neither $X_1$ nor $X_2$ contain any information about $X_3$; rather, information is completely stored in their joint state.

In a real-world example, the study of how two market indices cooperate to transfer information to a third one has shown non-trivial synergistic effects \cite{scagliarini2020synergistic} using the Partial Information Decomposition, a statistical framework that decomposes information in the unique, redundant and synergistic part but has the drawback of not being computationally feasible for more than three variables \cite{williams2010nonnegative}. A simple yet promising approach for quantifying high-order dependencies in a system is the O-information \cite{rosas2019quantifying}, which quantifies the balance between redundancy and synergy. Speaking qualitatively, in a complex system, the synergy measures its capacity to make integration of information whilst redundancy provides robustness to the system; decomposing interactions between variables into synergistic and redundant components illuminates how the system addresses the trade-off between robustness and integration. For example, recent works focusing on the brain at the macroscale have identified high synergy brain regions which support higher cognitive function \cite{luppi2022} and are affected by the aging process \cite{nuzzi2020}.

Many studies have investigated different aspects of the cryptocurrency time series, from their correlation properties \cite{aslanidis2019analysis,kruckeberg2019cryptocurrencies} to the possibility of being used for portfolio optimization \cite{briere2015virtual,elendner2018cross}.
The spillover effect between the major cryptocurrencies has been investigated by using VAR (Vector Autoregressive Models, as in \cite{koutmos_return_2018}, for instance), finding that return and volatility spillover is mainly driven by the Bitcoin and this effect was continuously rising during the considered period, suggesting a growing interdependence among cryptocurrencies. Other studies have investigated the relationship among the most important coins using VAR and SVAR (Structural VAR) Granger causality, finding that "Ethereum is likely to be the independent coin in this market, while Bitcoin tends to be the spillover effect recipient" \cite{luu_duc_huynh_spillover_2019}.
An example of direct use of Granger Causality test to study information flow between two major cryptocurrencies has been performed \cite{tu_effect_2019}, but this study was limited only to Bitcoin and Litecoin.
In \cite{garcia2020transfer}, multivariate transfer entropy, which is a model-free approach to study information flow \cite{schreiber2000measuring}, was used to study a network of cryptocurrencies, using a greedy algorithm to detect the most informative drivers. They found an increase in information flow during the market turbulence of March 2020. 
It has been shown that Granger Causality and transfer entropy are equivalent for Gaussian processes \cite{barnett2009granger}. Being  pairwise methods, neither Granger Causality nor Transfer entropy can assess high-order effects. To the best of our knowledge, there are no investigations among cryptocurrencies using high-order dependencies. The present work addresses to contribute in filling this gap, studying a high-order generalization of Granger Causality. We will show that the pairwise approach and the higher-order one give complementary information to describe the cryptocurrency trading network.

The paper is organized as follows. In section \ref{sec:data} and section \ref{sec:methods} we describe the dataset and the methodology used for the analyses, respectively. In section \ref{sec:results} we present and discuss the results while, in section \ref{sec:conclusions}, we close the paper drawing some conclusions.

\section{Data and preprocessing} \label{sec:data}

The dataset investigated in this study is freely available from Kraken crypto Exchange \cite{kraken} and contains traded prices in general between cryptocoins and currencies.

In the following we will use different names for cryptocurrencies according to their own properties. We refer more specific to a Coin to describe a cryptocoin with its blockchain, to a Token to describe a cryptocoin backed by a blockchain of another Coin. Also, we refer as Stablecoin to indicate to a Cryptocoin which has its price ``pegged'' to the price of another asset (usually the USD change value) and in general to a Fiat currency to any government-issued currency such as Euro, Dollar or others as described in \cite{britannica}.

Each file of the dataset is organized in ``pairs'', where one pair consist of a cryptocurrency traded using a fiat currency, so it reports all trade on a specific market, which collect bid and ask orders on a corresponding order book, from cryptocurrency to fiat currency and vice-versa. Each line refers to a single trade or transaction. Since it is possible to have more than one transaction per second referring to the same timestamp, with different price and volume, they are reported on separate lines.

For each pair, data provided for each trade have the following information:
\begin{itemize}
    \item {\emph{timestamp}} represents the time when the trade occurs, with resolution in seconds. It is possible to have more than one trade with the same timestamp, in this case volume and price may be different for each trade;
    \item{\emph{price}} is the price at which the cryptocurrency was traded, in terms of the corresponding fiat currency;
    \item{\emph{volume}} is the amount of cryptocurrency exchanged on the corresponding trade;
\end{itemize}

Data were aggregated on the scale of minutes, considering the weighted mean over volume for the minute price of the trade, and the sum of volumes for the total volume exchanged at the current minute time. If there were no trade on a minute, in order to avoid discontinued time series, we hold the previous last price to fill the gap on the time series, setting the corresponding volume to zero.
Then, we computed the logarithmic returns of each time series of the price, which will be divided into time series of weekly length, to see how the results vary over time and to take into account effect due to the different volume of transactions over the week (usually smaller volumes are traded during weekends).
We divided the considered period in weekly time windows, each starting at 00:00:00 every Monday and terminating at 23:59:59 every Sunday (with the exception of last window because last day available on the dataset was Friday, 31 Dec 2021), thus obtaining $104$ windows, each containing a total of 10080 points.

We remark that the analysis is performed just on the time series of logarithmic returns; the time series of volumes are used here only to compare the time evolution of information-theoretical quantities with the total volume of exchanges in US dollars. A cryptocurrency can be traded on the Exchange using more than one fiat, then there is the possibility to have more than one pair for each crypto, one for each available fiat market (i.e. Bitcoin can be traded in EUR, Dollar or other fiat currencies related to the country of the user). Also, the Exchange allowed to use some crypto coins (i.e. Bitcoin, Ethereum) as fiat, probably to speed up buying and selling operations.

Table ~\ref{tab:kraken} in Appendix \ref{app:tables} reports a brief description of all data available on Kraken, grouped by years, number of cryptos and fiat, while table ~\ref{tab:coin_description} shows which tickers are included on the dataset and some of their features. 

As the cryptocoin market grows, the number of cryptos and fiat listed on Kraken increase to about hundred cryptocoins, traded up to 12 different market, for a total of about 450 crypto-fiat pairs.

In our analysis, we focus on the period 2020-2021 since the number of cryptos is large and the volume of the transaction is also large. It is possible to have an overview of the data related to each cryptos for 2020-2021 on table \ref{tab:data_coverage}.

\section{Methods} \label{sec:methods}
\subsection{Pairwise Granger Causality} \label{pairwise_granger_causality}
In this section, we describe how to implement Granger Causality between pair of variables in the context of linear autoregressive models.

Let us consider two stationary time series $X_{t}$ and $Y_{t}$. 
Under suitable conditions \cite{hamilton2020time}, the target series $Y$ can be written as a weighted sum of its past states and an error term $\epsilon'_t$ (reduced model); similarly, a second model can be built by adding the past states of the source variable (full model):
\begin{align}
    Y_{t} &= \sum_{m=1}^p a_{m}Y_{t-m} + \epsilon'_t \\
    Y_{t} &= \sum_{m=1}^p a_{m}Y_{t-m} + \sum_{m=1}^q b_{m}X_{t-m} + \epsilon_t~
\end{align}
In other words, in the full model, the future states of the target are predicted using its previous $p$ past states and the previous $q$ states of the source, while in the reduced model only the target is involved.

The goodness of these models can be quantified by the variance of the error term of the full model $\sigma^2_f = \mean{\epsilon^2_t}$ and of the reduced model $\sigma_{r}^2 = \mean {\epsilon'_t}$, where $\mean{\cdot}$ indicates the expected value. To test if the full model improves the predictability of the target, we use the statistics 

\begin{equation}\label{eq:gc_pair}
    \mathcal{F}_{X\rightarrow Y} = \ln \frac{\sigma_{r}^2}{\sigma^2_f}~.
\end{equation}
This choice has a number of useful properties \cite{geweke_measurement_1982}, remarkably the invariance under scaling transformation of the original time series and the asymptotic chi-square distribution under the null hypothesis $\mathcal{F}_{X\rightarrow Y} =0$. To build a bridge between Granger causality and the information-theoretical methods for analyzing high-order dependencies of the next section, we can consider the equivalence between Granger causality and transfer entropy $\mathcal{T}$ under the Gaussian approximation \cite{barnett2009granger}. 

Transfer entropy is a model-free version of Granger causality. Considering the source states $\bm{X}^{-}_t = (X_{t-1},X_{t-2},\ldots,X_{t-p})$, at varying $t$, as realizations of a stochastic variable  $\bm{X}^-$, and analogously $\bm{Y}_t^-= (Y_{t-1},Y_{t-2},\ldots,Y_{t-p})$ as realizations of the target state variable $\bm{Y}^-$,  $Y_{t}$ as realizations of the future of the target y, considered as a stochastic variable, the transfer entropy is defined as the mutual information between source and target variable conditioned over the target past states:
\begin{equation}
   \mathcal{T}_{X\rightarrow Y} = I(y ; \bm{X}^-|\bm{Y}^-)
\end{equation}
The equivalence reads: 
\begin{equation}
    \mathcal{F}_{X\rightarrow Y} = 2 \mathcal{T}_{X\rightarrow Y} 
\end{equation}

The pairwise Granger causality has been computed for each pair using the MATLAB toolbox MVGC \cite{barnett_mvgc_2014}.
The result of Granger causality produce an adjacency matrix $A = \{a_{ij}\} $, where the element $a_{ij}$ represents the Granger Causality from the cryptocoin $i$ toward the cryptocoin $j$, which can be used as weighted links for our directed Cryptocurrencies Trading Network.

To find the most influencing and influenced nodes, one may define the quantities out strength and in strength:

\begin{equation}
\label{outstrength}
    k^{{\rm out}}_i = \sum_{j\neq i} a_{ij}
\end{equation}

\begin{equation}
\label{instrength}
    k^{{\rm in}}_i = \sum_{j\neq i} a_{ji}
\end{equation}

\subsection{O-information} 
\label{sec:oinformation}
We now describe how to study high-order dependencies in a  collection of random variables $\bm{X}^n = (X_1,X_2,\ldots,X_n)$. 
The first step is to extend the concept of mutual information to the multivariate case: a popular way to do it is through \emph{total correlation} (also known as multi-information)\cite{watanabe_information_1960}, which quantifies the amount of high-order constraints and then is a measure of redundant information. In terms of information-theoretical quantities, it can be expressed as:
\begin{equation}
    \TC(\bm{X}^n) = \sum_{k=1}^n H(X_k) - H(\bm{X}^n),
\end{equation}
where $H(X_k)$ and  $H(\bm{X}^n)$ are the Shannon entropies, which quantify the uncertainty of the k-th variable and of the whole system, respectively. 
Another useful way of looking at this quantity is as a measure of the "statistical constraints" acting on $X_k$: when $H$ is low, the system explores more frequently small regions of the phase space and, in this sense, the constraints are high. Conversely, when $H$ is maximum (a condition that occurs when the distribution is uniform), then there is the minimum possible amount of constraints acting on $X_k$.

Another popular extension of mutual information is the \emph{dual total correlation} \cite{sun_linear_1975}
\begin{equation}
    \DTC(\bm{X}^n) = H(\bm{X}^n) - \sum_{k=1}^n H(X_k\vert \bm{X}^n_{-k})
\end{equation}
where $\bm{X}^n_{-k}$ represents all the system minus the k-th variable. $\DTC(\bm{X}^n)$  may be interpreted as the amount of uncertainty that can be explained only by observing more than one variable at once: for this reason, dual total correlation is a measure of the synergy present in the system at global level.
Crucially, it can be easily shown that both $\TC(\bm{X}^n)$ and $\DTC(\bm{X}^n)$ are nonnegative \cite{scagliarini_quantifying_2022} and then can be used as proper measures of redundancy and synergy.

The following quantity, called O-information \cite{rosas2019quantifying},  represents the balance between the redundant and the synergistic dependencies in $\bm{X}^n$:
\begin{align}
    \Omega(\bm{X}^n) &= \TC(\bm{X}^n) - \DTC(\bm{X}^n) \\ 
                &=(n-2)H(\bm{X}^n)+ \sum_{k=1}^n       \big[H(X_k) - H(\bm{X}^n_{-k}) \big].
\end{align}
Accordingly, if this quantity  is greater than $0$, we say that the system is redundancy-dominated, whilst if $\Omega < 0$ the system is synergy-dominated.

Now, we introduce a target variable $y$ to study whether the high-order dependencies between a multiplet of $n$ variables $\bm{X}^n$ with $y$ are redundant or synergistic. 
We can express the O-information of the system jointly formed by the multiplet and the target as follows
\begin{equation}
    \Omega(\bm{X}^n\cup y) = \Omega(\bm{X}^n) + \Delta^y
\end{equation}
where $\Delta^y$ the contribution  of a variable to the O-information and reads: 
\begin{equation}
\Delta^y = (1-n)I(y;\bm{X}^n) + \sum_{k=1}^n I(y;\bm{X}^n_{-k}),
\end{equation}
where $I(y;\bm{X}^n) = H(y) + H(\bm{X}^n) - H(y;\bm{X}^n)$ denotes mutual information.

Now we turn to consider $n$ time series $\{X_{i,t}\}$ and a target  $\{Y_{t}\}$, where $t=1,\ldots,T$ and $i=1,\ldots,n$; 
the informational character of the information flow from  the multiplet of X variables (considered as sources) to the target, can be assessed introducing stochastic variables $\bm{X}^{n-} = (X_1^-,X_2^-,\ldots,X_n^-)$,  $\bm{Y}^{-}$ and $y$ representing the sources, the target and the future of the target respectively, whose realizations can be obtained from the samples multivariate time series.

Hence, the dynamical version of the O-information, introduced in \cite{stramaglia_quantifying_2021} for the analysis of neural signals, reads
\begin{equation}
         d\Omega^y(\bm{X}^n) = (1-n)I(y;\bm{X}^{n-}|\bm{Y}^-) + \sum_{k=1}^n I(y;\bm{X}^{n-}_{-k}|\bm{Y}^-).
\end{equation}

This quantity can be also thought as a high-order version of the Granger causality and its sign has the same interpretation of the O-information in terms of redundancy and synergy.

In order to analyse the dataset, we proceed as follows. First, we fixed a target variable $y$; then, we searched among the remaining $n-1$ variables for the pair that maximize and minimize the quantity
$$
d\Omega^y_{\text{red}}(\bm{X}^2) = \max_{i,j} d\Omega^y(\{X_i X_j\})
$$
$$
d\Omega^y_{\text{syn}}(\bm{X}^2) = \min_{i,j} d\Omega^y(\{X_i X_j\})
$$
Then, $d\Omega^y_{\text{red}}(\bm{X}^2)$ and $d\Omega^y_{\text{syn}}(\bm{X}^2)$ is the highest and lowest dynamic O-information to the target $y$ from the best multiplets ${\bm \tilde{X}^2} = \{X_{i^*} X_{j^*}\}$ of size $n=2$. For sizes $n>2$,  we used a greedy approach: for $n=3$, we start from the ${\bm \tilde{X}^2} = \{X_{i^*} X_{j^*}\}$ found in the previous step, and for the remaining variables $X_k$ we choose the one that maximizes and minimize $d\Omega^y(\{X_i X_j X_k\})$. We repeated the same procedure for $n=4$ and so on,  and we stop at $n=5$.

\section{Results and Discussion} \label{sec:results}

In this section we present and discuss the results obtained with both the pairwise Granger causality approach and the O-information methodology, evaluated among the logarithmic returns time series of cryptocurrencies exchanged in US dollars (USD) on the same time window. 

\subsection{Pairwise Granger Causality}

For each one of the 104 weekly windows (from 30 Dec, 2019 to 31 Dec, 2021), Granger causality (GC) analysis was conducted using Equation \eqref{eq:gc_pair} for all the $n(n-1)/2$ pair of variables active in that window, which become nodes of the cryptocurrency traded network. For each pair, the optimum model order $p$ was chosen using the Bayesian information criterion \cite{schwarz_estimating_1978}. A link between a pair of nodes is created only if the GC strength overcomes the significance threshold of $1\%$. 

\begin{figure}[!ht]
    \centering
    \includegraphics[width=0.8\textwidth]{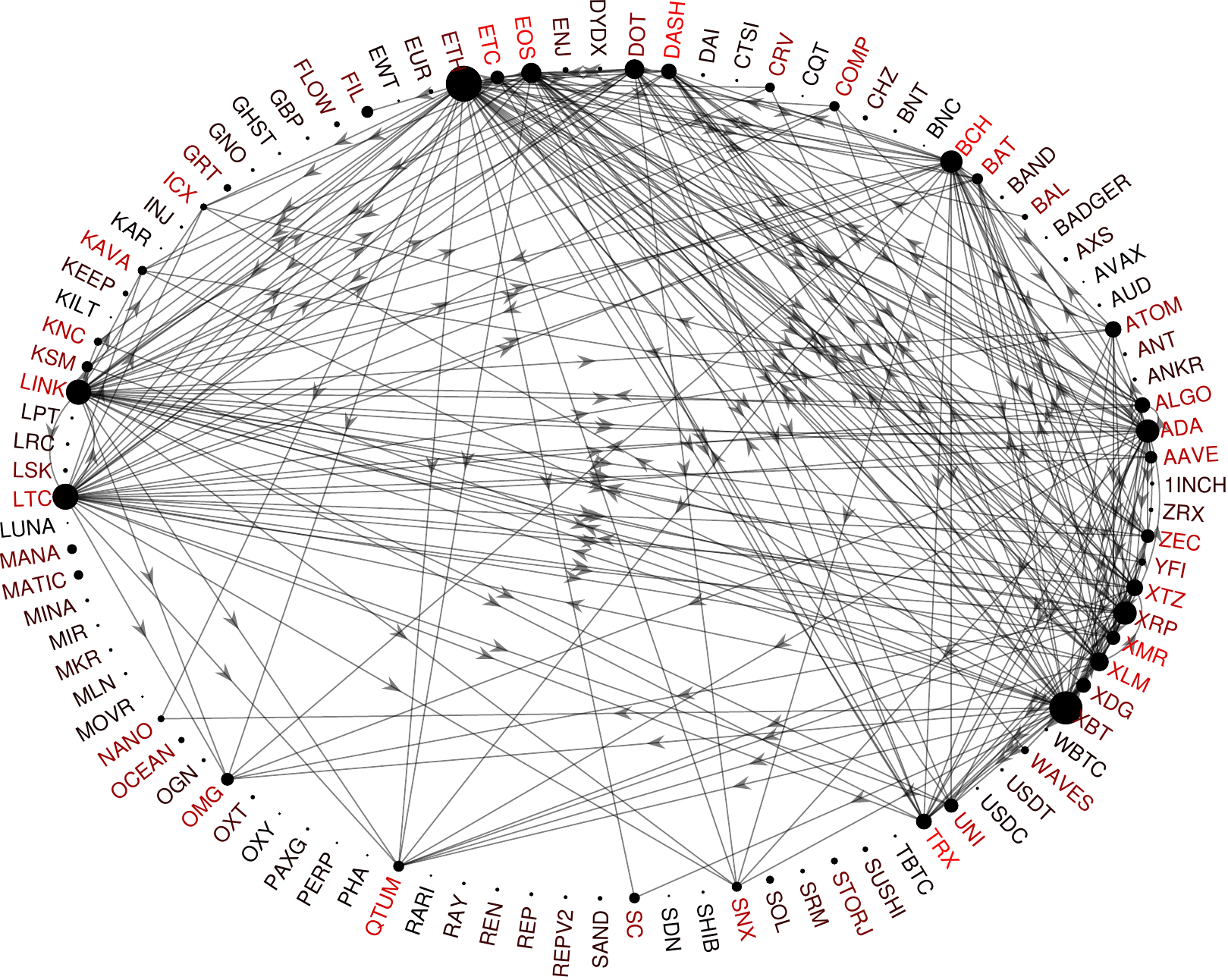}
    \caption{{\bf Average USD network.} Each link $\overline{a}_{ij}$ represents the Granger Causality averaged over the 104 weekly windows from 28 Dec, 2019 to 30 Dec, 2021. Node size represents the total out-strength $k^{{\rm out}}_i = \sum_j \overline{a}_{ij}$, while the label color (from black to red) is the total in-strength $k^{{\rm in}}_i = \sum_j \overline{a}_{ij}$, that is how much that node is influenced by the rest of the network. }
    \label{fig:network_medio}
\end{figure}

A first visual representation of the structure of the USD network averaged over all the time windows is shown in Figure \ref{fig:network_medio}, where the biggest nodes represent the most influential elements and the color scale of the labels indicates how much a node is influenced by all the others (for a better visualization, only the first $5\%$ of links sorted according to their GC strength is reported in the figure). This kind of graph is very effective in giving an overall idea about the relative importance of cryptocurrencies in the whole considered period (the network evolution across weekly time windows will be addressed later).

\begin{figure}[!ht]
    \centering
    \includegraphics[width=0.9\textwidth]{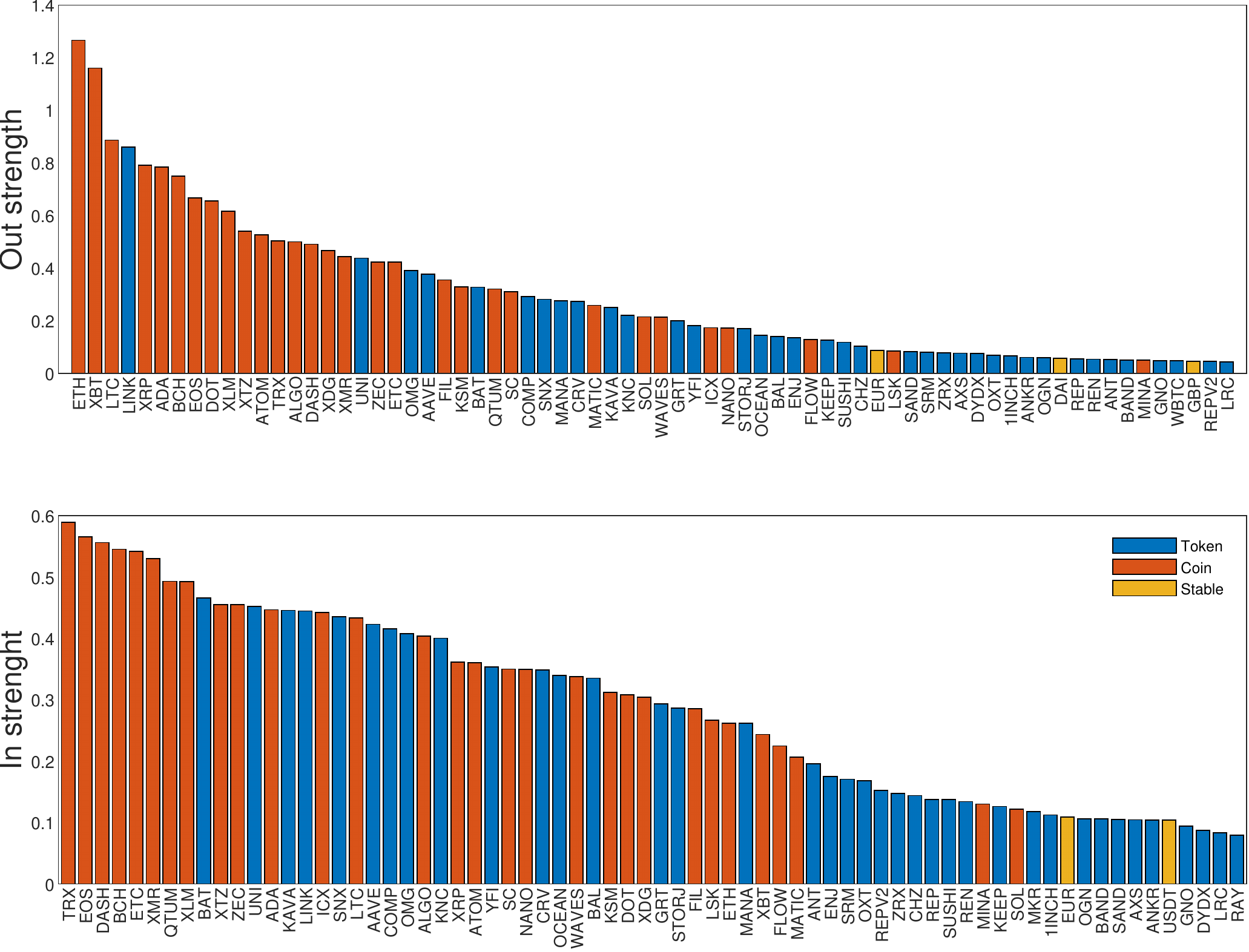}
    \caption{{\bf  The most influential and influenced nodes in the USD network of pairwise Granger causality.} On the y-axis is indicated the total in-strenght and out-strength . Yellow bars indicates stable-coins, red bars coins and blue bars tokens. Only the first $70$ elements are shown.}
    \label{fig:bar_plot_granger}
\end{figure}

To see more in detail which cryptocurrencies transfer the most information to the rest of the system and which receive the most information from the others, it is useful to look at the ranking of the average in strength and out strength, calculated with Equation \eqref{outstrength}. These information are shown, respectively, in the top and bottom panels of Figure  \ref{fig:bar_plot_granger}. The different classes of cryptocurrencies - token, coin and stable coins - are represented using different colors.
As can be seen in the top panel, the most influential cryptocurrencies belong to the class ``coin" (the ones that have their own blockchain, colored in red). Unsurprisingly, the most influential coins are the eldest and most capitalized, such as Bitcoin (XBT), Ethereum (ETH), and Litecoin (LTC). Among the topmost influenced elements (bottom) we find younger and less capitalized ``coins", such as Tron (TRX), EOSIO (EOS) and Dash (DASH). Stable coins seem to play a marginal role in the pairwise description of the network, both for transferred (EUR, DAI, GBP) and received information (EUR, USD).

\begin{figure}[!ht]
\centering
\begin{tabular}{cccc}
\subcaptionbox{28-Dec-2020 \label{1}}{\includegraphics[width = 2in]{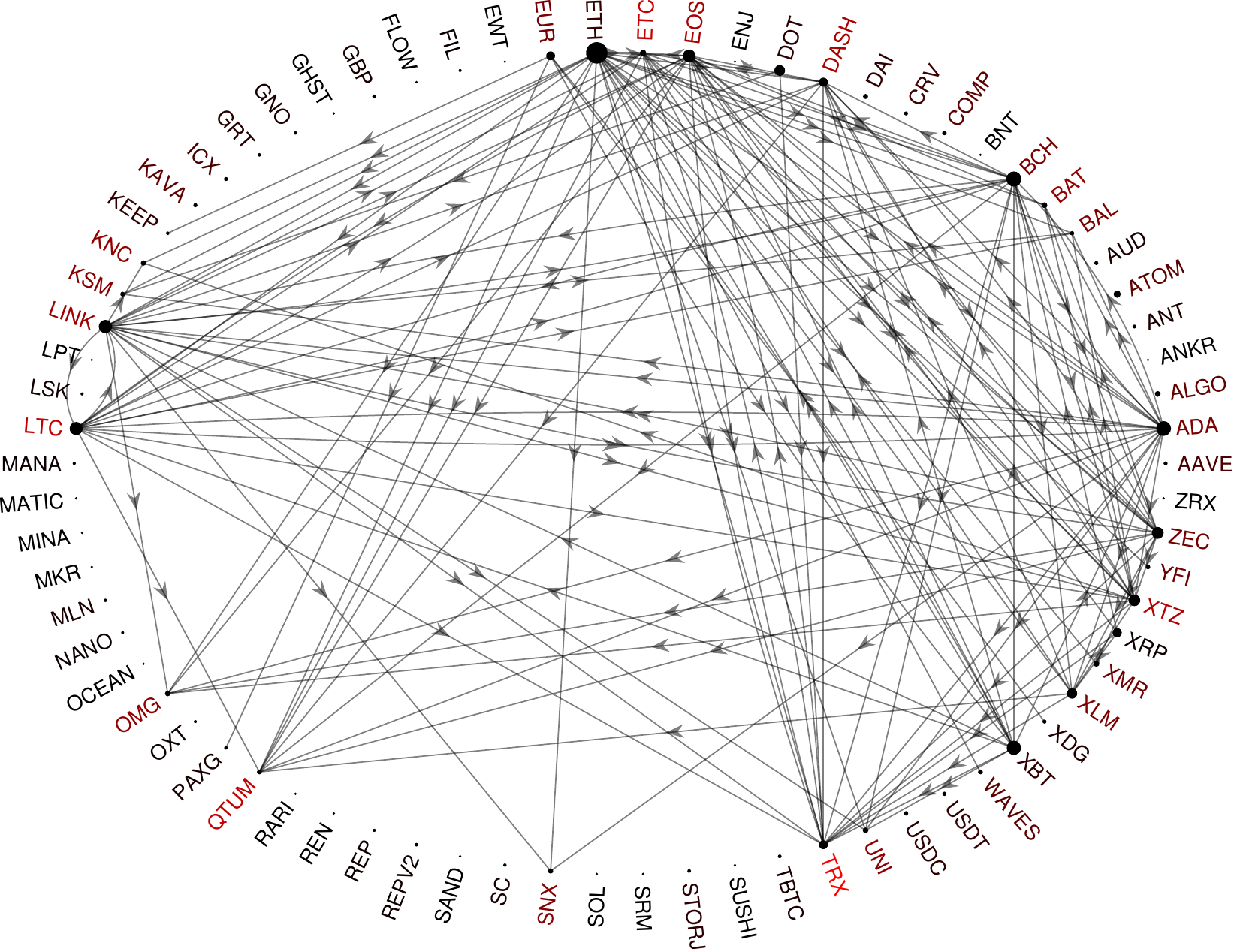}} &
\subcaptionbox{04-Jan-2021}{\includegraphics[width = 2in]{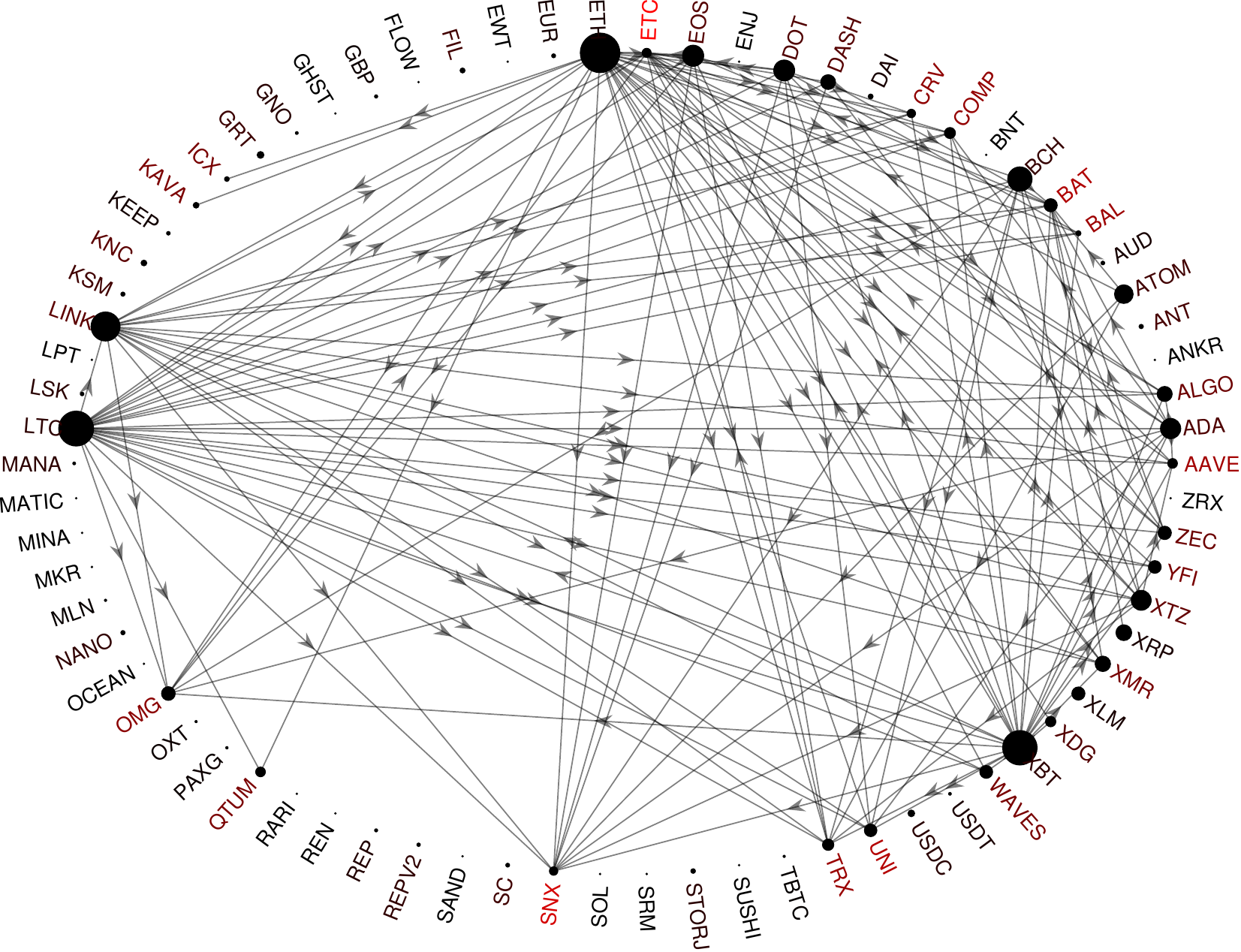}} &
\subcaptionbox{11-Jan-2021}{\includegraphics[width = 2in]{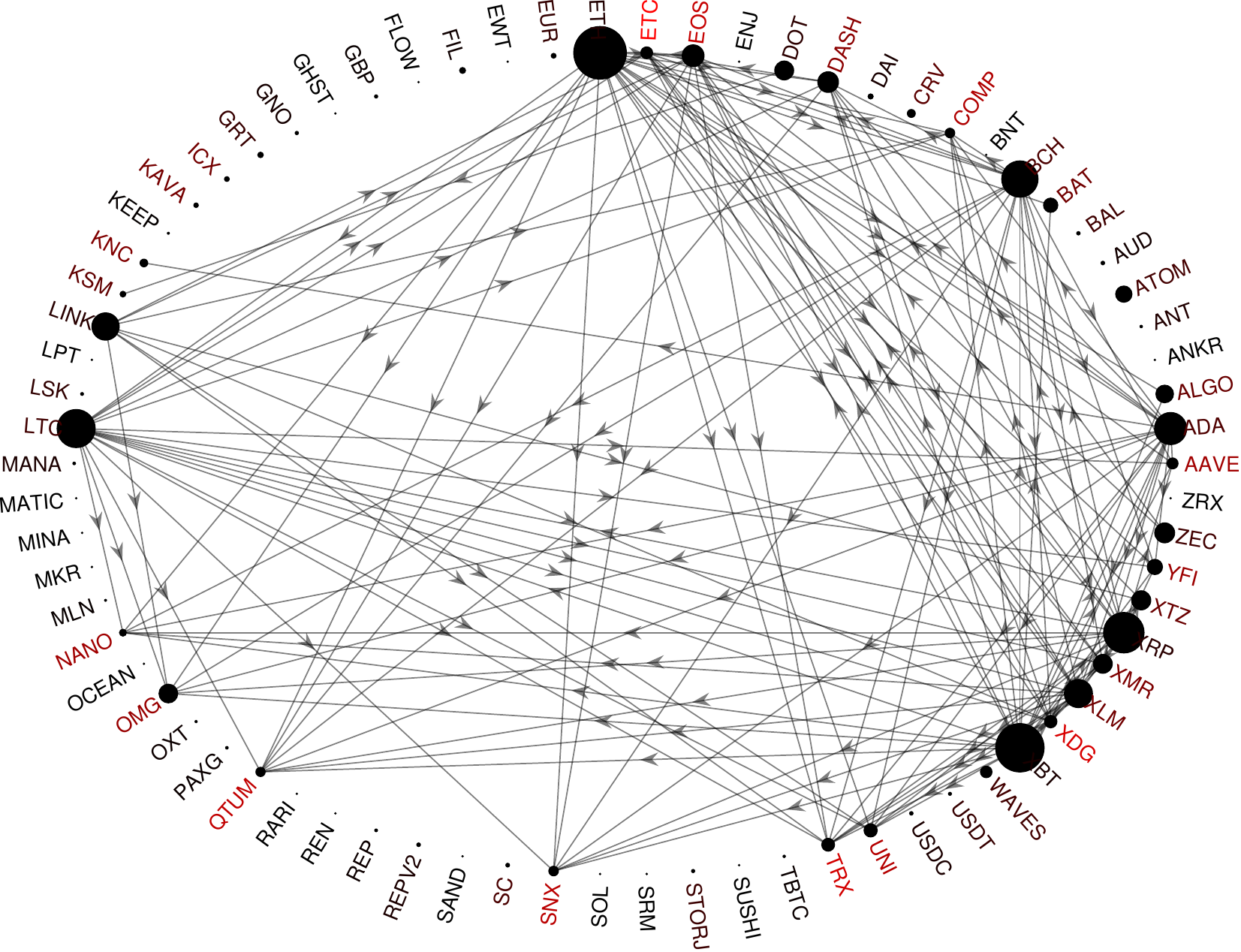}} \\\\
\subcaptionbox{18-Jan-2021}{\includegraphics[width = 2in]{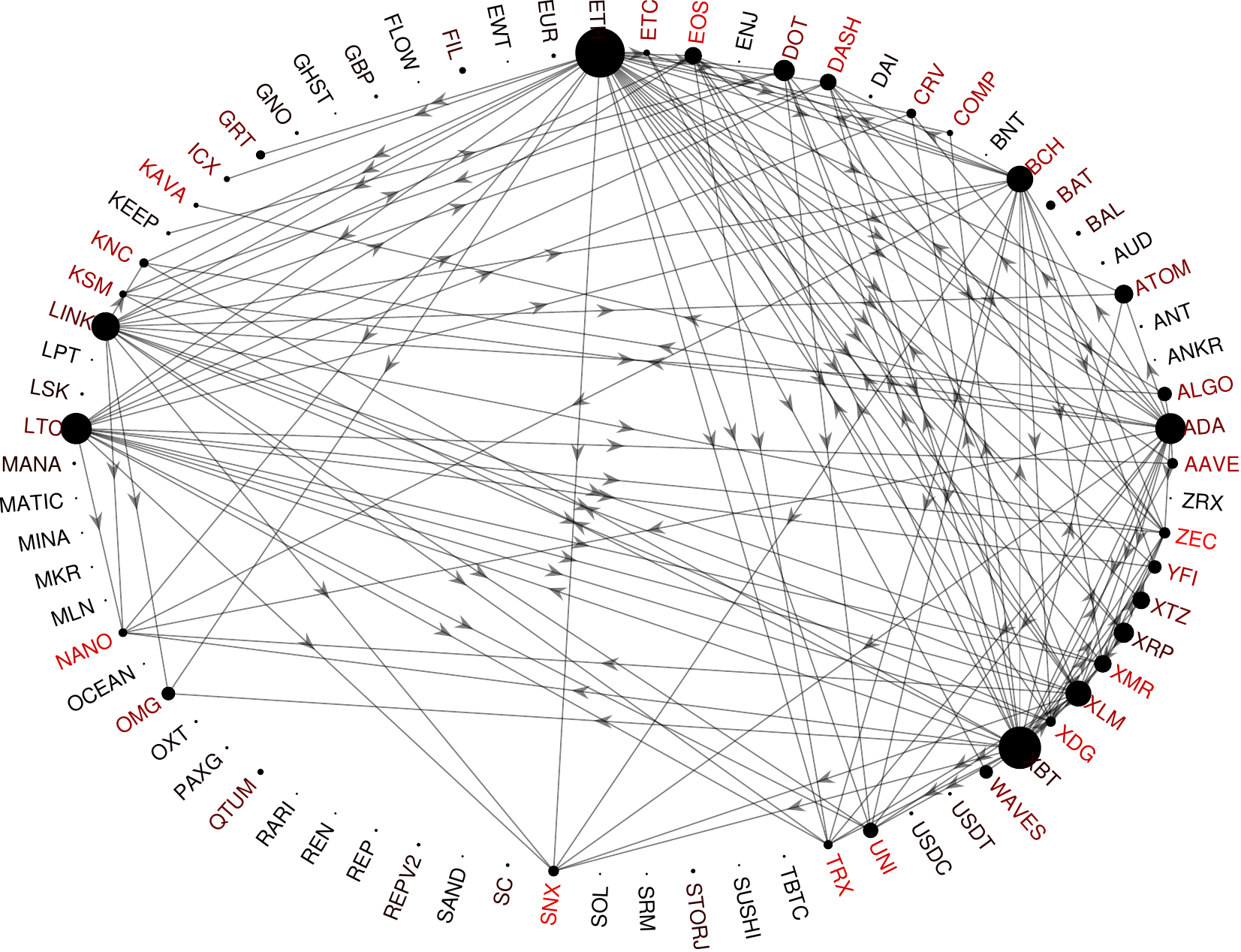}} &
\subcaptionbox{25-Jan-2021}{\includegraphics[width = 2in]{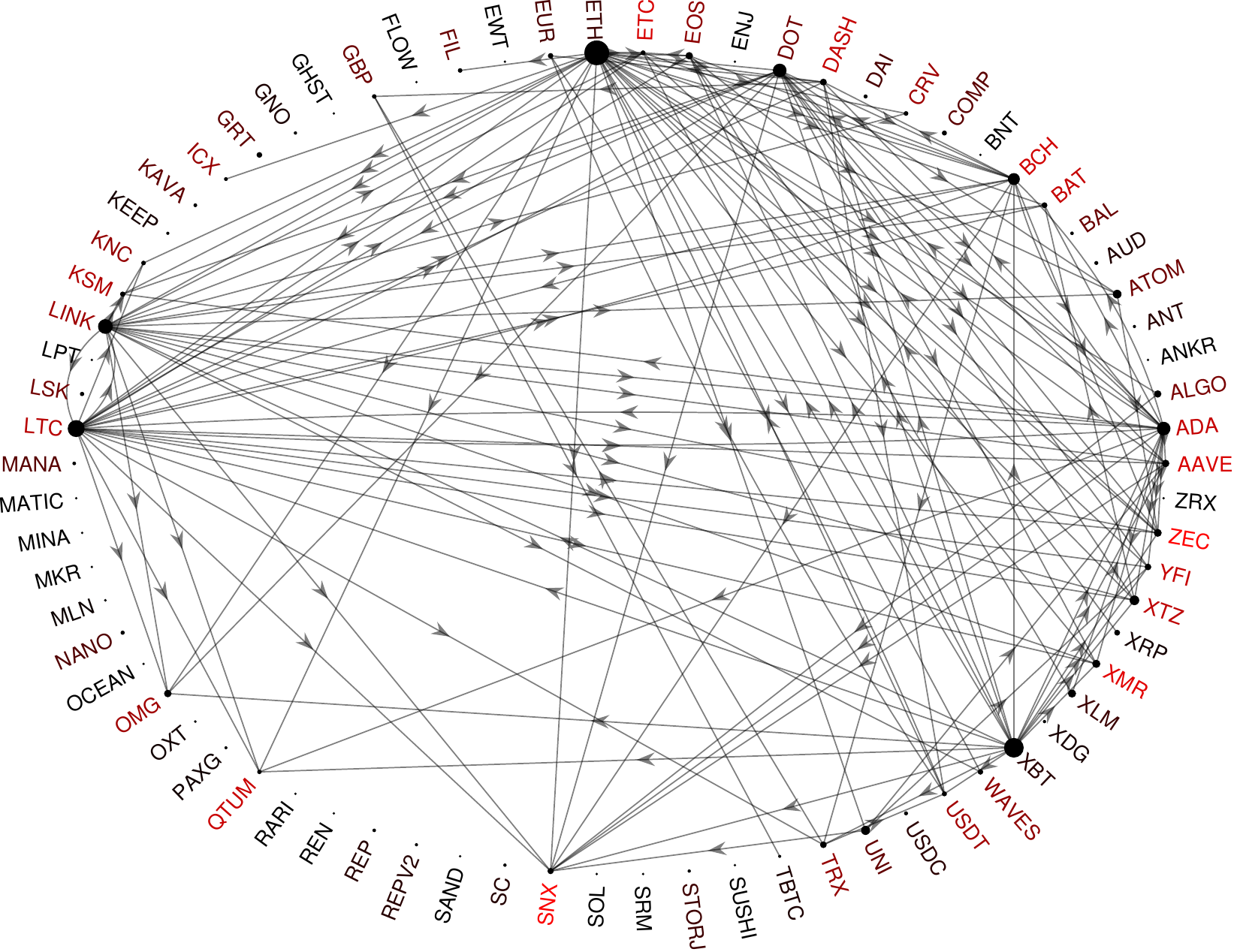}} &
\subcaptionbox{01-Feb-2021}{\includegraphics[width = 2in]{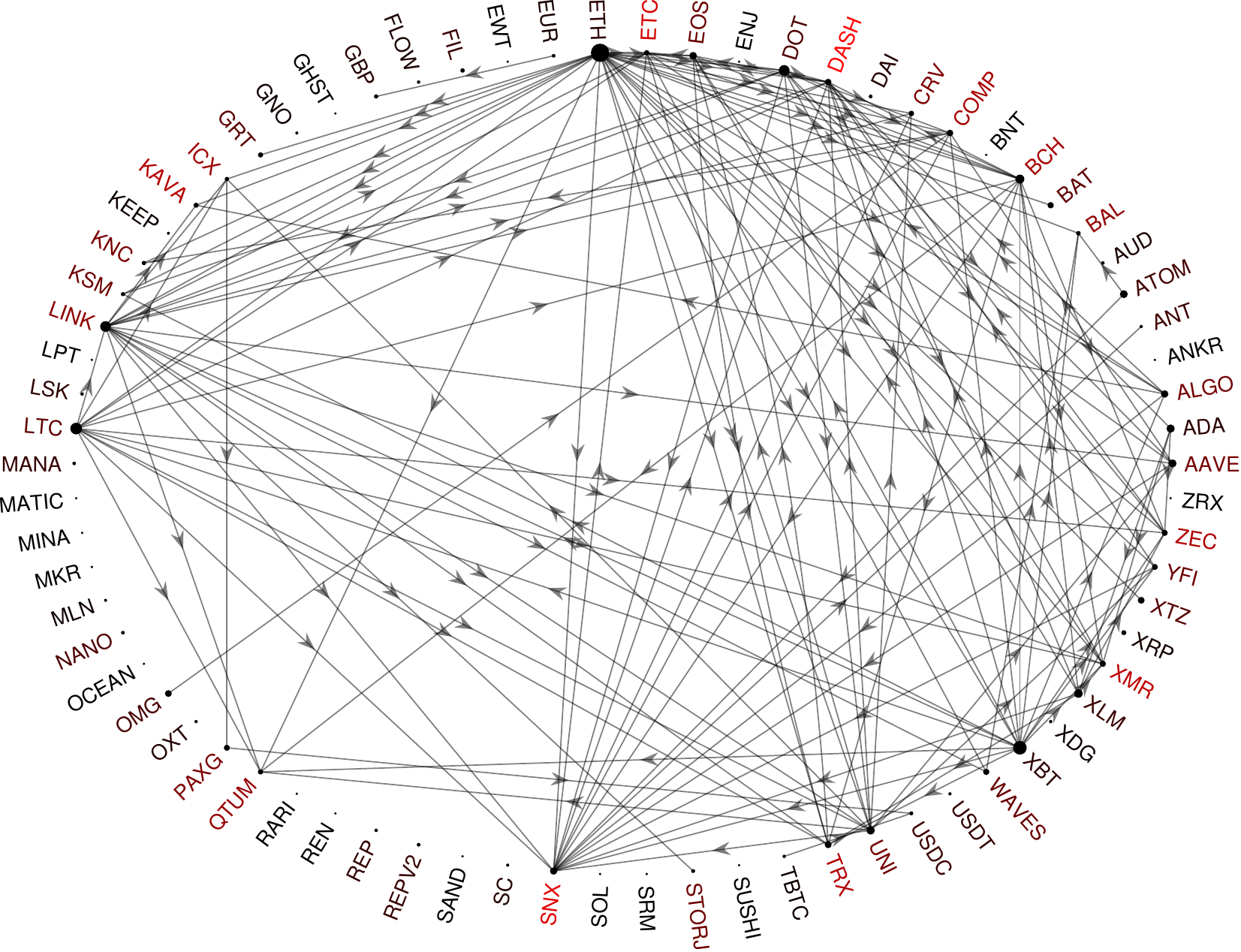}} \\\\
\subcaptionbox{08-Feb-2021}{\includegraphics[width = 2in]{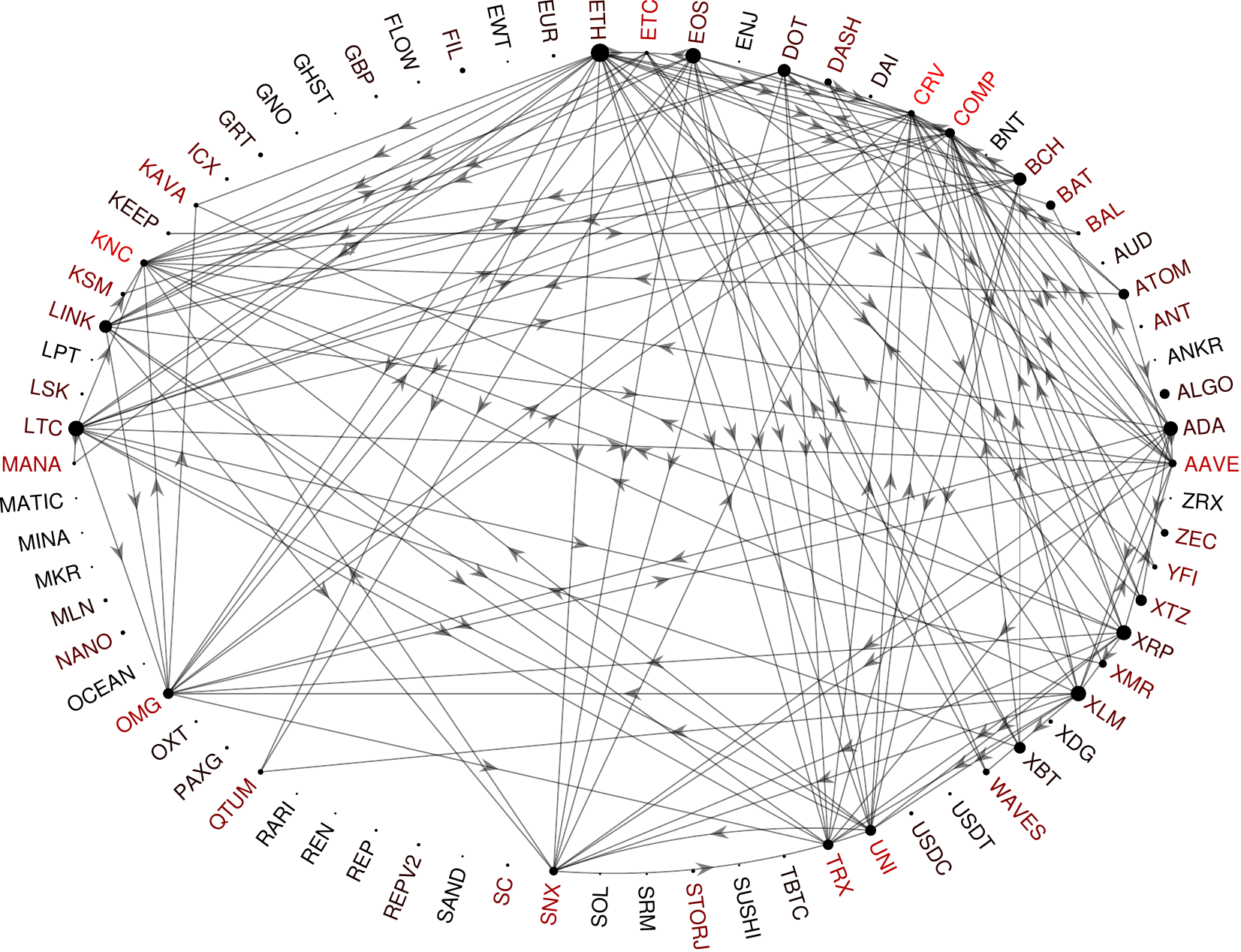}} &
\subcaptionbox{14-Feb-2021}{\includegraphics[width = 2in]{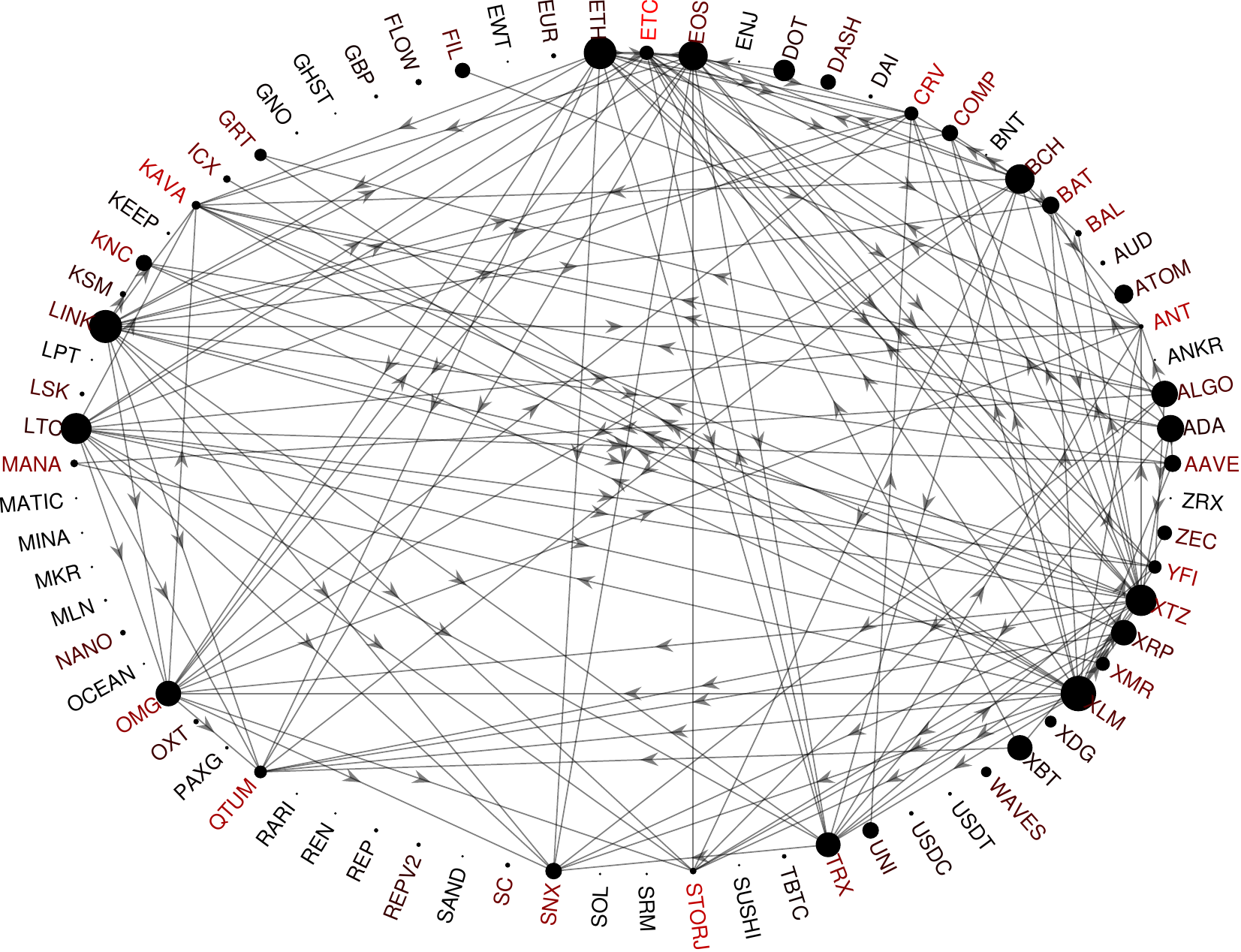}} &
\subcaptionbox{22-Feb-2021}{\includegraphics[width = 2in]{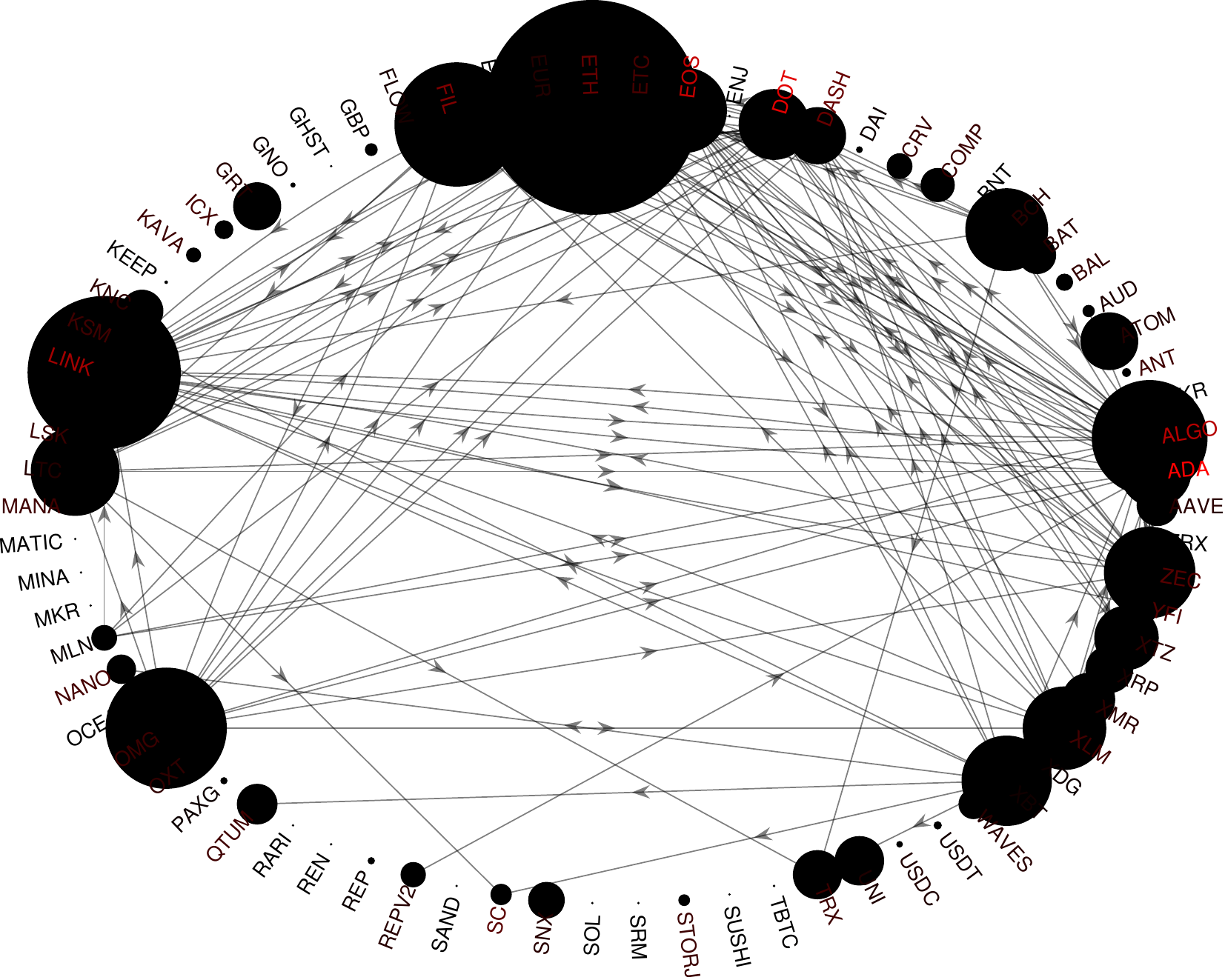}} \\\\
\subcaptionbox{01-Mar-2021}{\includegraphics[width = 2in]{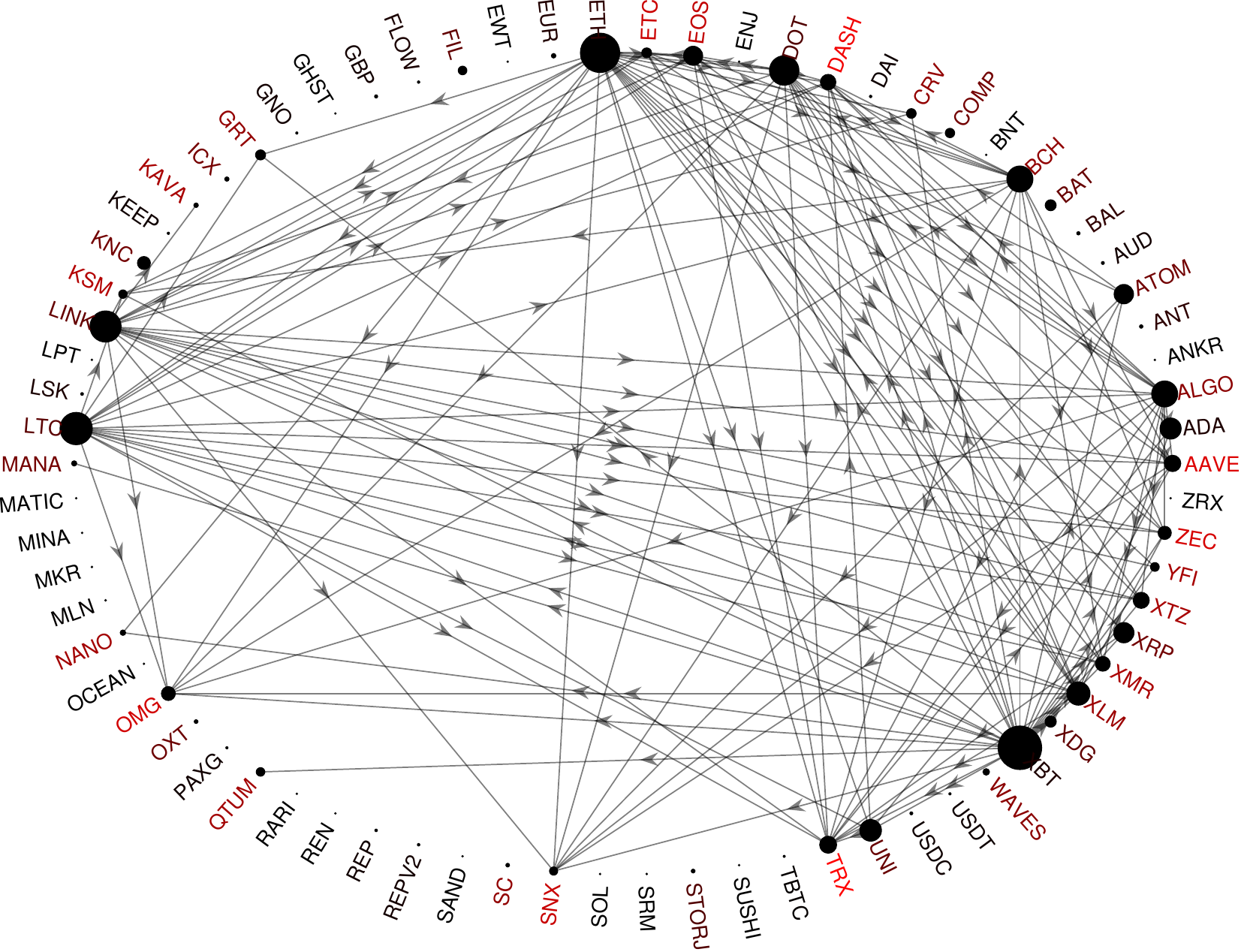}} &
\subcaptionbox{08-Mar-2021}{\includegraphics[width = 2in]{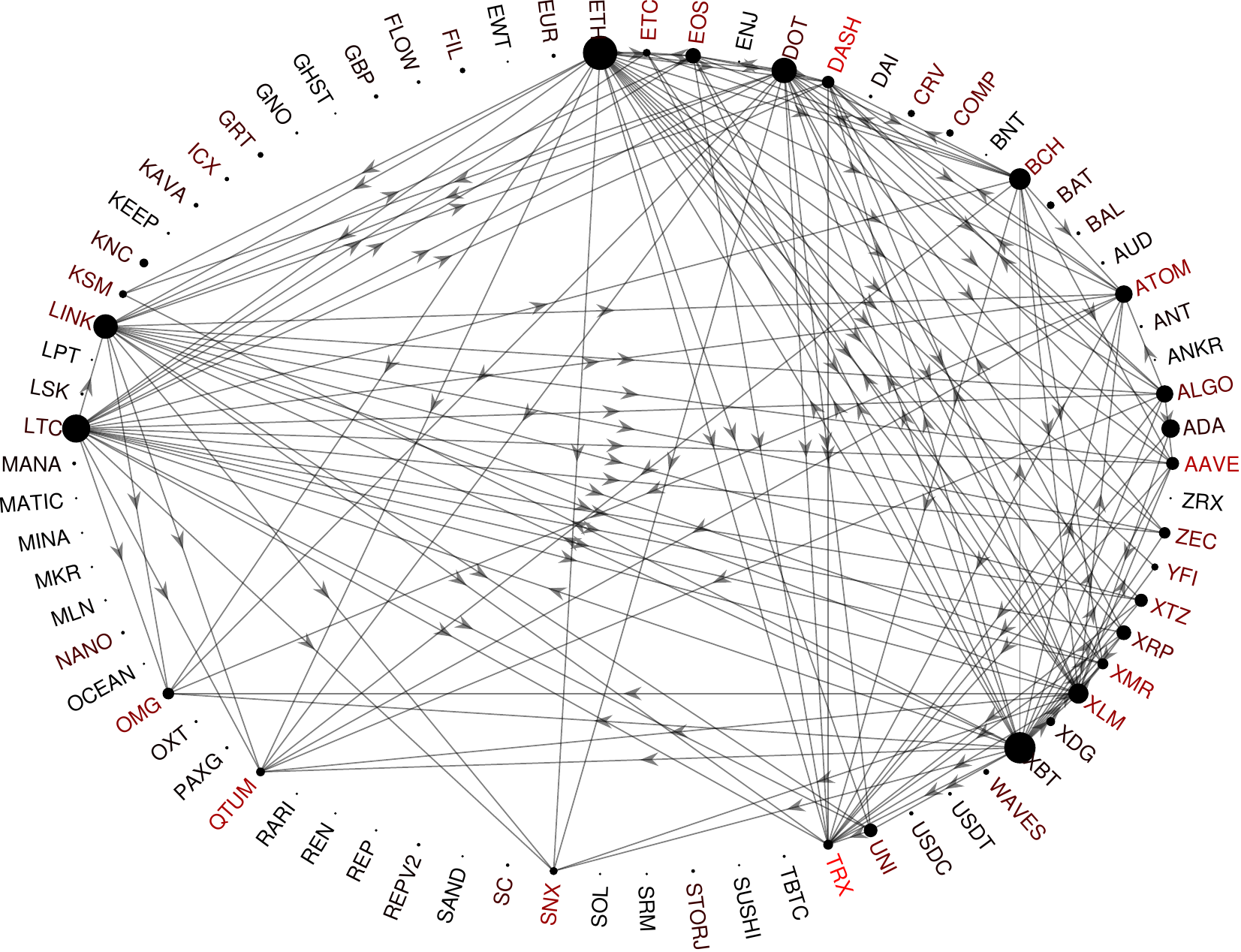}} &
\subcaptionbox{15-Mar-2021}{\includegraphics[width = 2in]{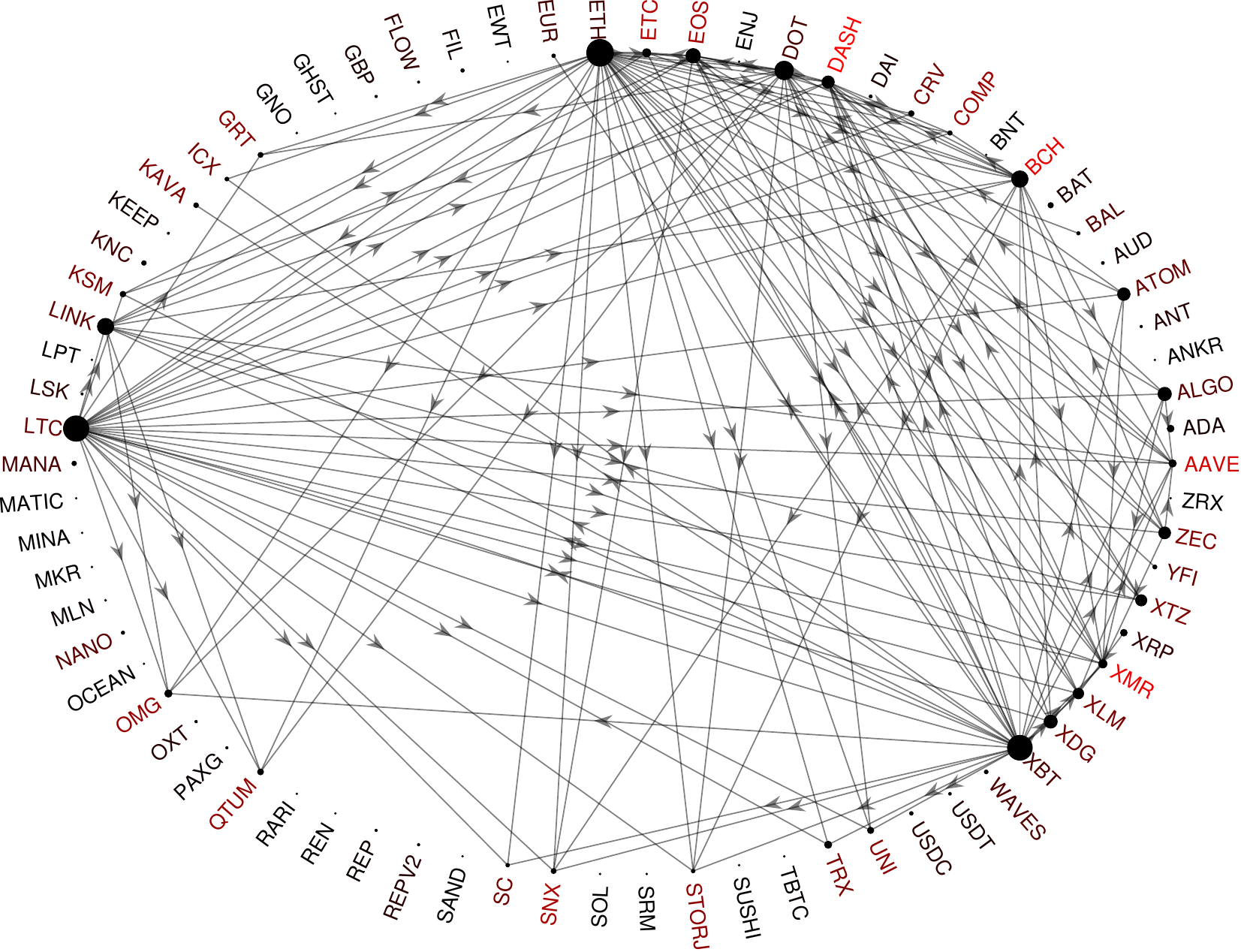}}
\end{tabular}
\caption{{\bf The USD network of pairwise Granger causality for weekly time windows.} Node size represents the total out strength of that node or, in other words, how much that node influences the network, while the label color (from black to red) is the total in strength, that is how much that node is influenced by the rest of the network.}
\label{fig:networks}
\end{figure}

To get a visual hint of how the information flow changes over time, in Figure \ref{fig:networks} we depict the structure of the USD network in several weekly windows (for a period going from 28 Dec, 2020 to 22 Mar, 2021), selected for being stable windows or characterized by the presence of crashes or turbulence. For example, an anomalous activity involving several currencies with high out strength  (biggest nodes) can be appreciated in the week starting on 22 Feb, 2021, while during the majority of the other weeks the information flow is visibly much lower. It is worth to notice that the number of nodes in the network slightly increases for subsequent windows, since new cryptocurrencies enter in the market/dataset.      

In this regard, it is interesting to investigate if there is some correlation between the age of the cryptocurrencies (i.e. the number of time windows in the USD dataset where it is present) and their in and out strength.  As can be seen in the first row of Table \ref{tab:birth_corr}, actually it seems that such a correlation does exist and is also quite consistent (expecially for the out strength). 

\begin{table}[!h]
    \centering
    \normalsize
    \begin{filecontents*}{./figures/birth_corr.csv}
fiat   , in strength, out strength
USD, $0.304^{**}$, $0.442^{**}$
EUR, $0.246^{*}$, $0.400^{**}$
XBT, $0.015$, $0.051$
ETH, $-0.152$,$-0.076$
\end{filecontents*}
\csvreader[tabular=c|c|c, table head=\toprule, table foot=\bottomrule, no head,
    late after line=\\, late after first line=\\\midrule,]{./figures/birth_corr.csv}{}{\csvcoli & \csvcolii & \csvcoliii}
    \caption{Table shows the Spearman correlation $\rho$ between the in-strength and the out-strength of a fiat with its ``age" (that is, the number of windows in which it was active in the dataset). Two asterisks indicate that $\rho$ is significant at $p<0.01$, one asterisk at $p<0.05$. }
    \label{tab:birth_corr}        
\end{table}

\setlength{\tabcolsep}{6pt}

Repeating the same procedure also for the other datasets (where the cryptos are exchanged in different fiats, i.e. EUR, Bitcoin or Ethereum), it results that the correlation stays quite high only when EUR is adopted as fiat (second row in Table \ref{tab:birth_corr}) while it rapidly disappear for the other fiat currencies (third and fourth rows). 
This can be explained by the fact that when a given fiat is firstly adopted for the exchange, it is still quite unpopular, and then its connections with the rest of the system from the point of view of Granger causality are still weak. Conversely, as time passed, its importance grows and this is reflected in the values of information flow.

\begin{figure}[!h]
    \centering
    \includegraphics[width=0.70\textwidth]{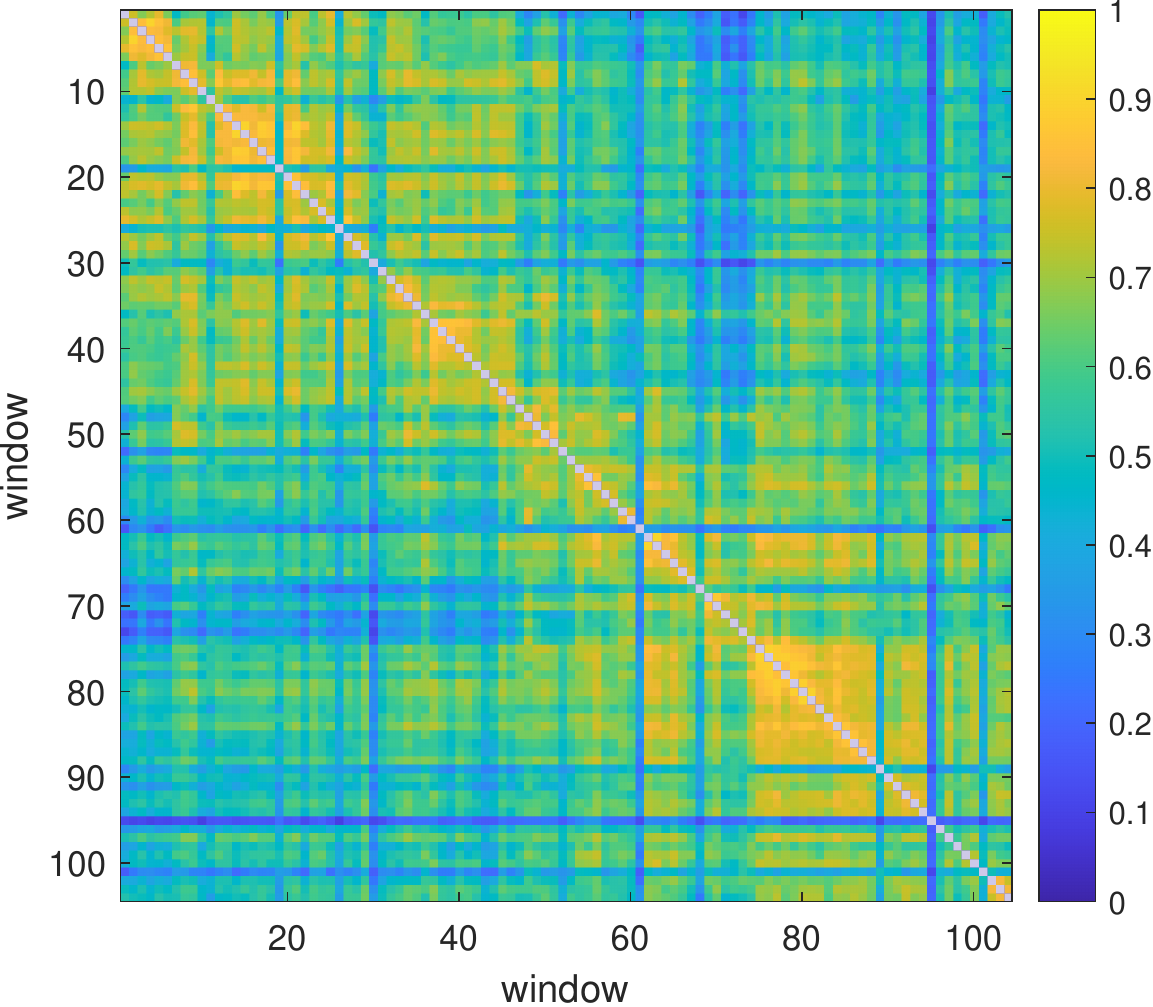}
    \caption{{\bf The Pearson correlation between adjacency matrices at different time windows.}
The element $c_{ij} = \rho(A^i,A^j)$  of the matrix is the Pearson correlation between the vectorized adjacency matrices at windows $i$ and $j$. High values of $c_{ij}$ indicate that the structure of the network at windows $i$ and $j$ is very similar.
Diagonal elements, indicating self-correlation, are colored in gray.}
\label{fig:corr_2020_2021}
\end{figure}

Finally, it is also interesting to look at the correlation between pair of USD networks of Granger causality as they appear in any couple of windows $h$ and $k$. To this aim, in Figure \ref{fig:corr_2020_2021} we report the Pearson correlations $\rho(A^{h},A^{k})$ between the adjacency matrices $A^{h}$ and $A^{k}$, for each $h,k=1,2,\ldots,104$. 
Correlation $\rho$ can be seen as a measure of similarity between networks: high values of $\rho$ in two different windows indicate that, on average, the information flows between the cryptocurrencies are similar in the corresponding two periods. If we look at values close to the main diagonal, we see that for consecutively windows the correlation is very high, often above $0.9$; also, inspecting windows more separate in time (that is, looking at values far from the main diagonal) correlation still remains quite high, indicating that the structure of the Granger causality network is quite stable or, in other words, that the information flow between the various nodes does not change much over time.

Analogous correlation matrices calculated for the other datasets (EUR, XBT and ETH) are reported for completeness in Figure \ref{fig:corr_2020_2021_all} of Appendix \ref{app:otherfiat} (of course, as expected for the reasons explained above, correlations between networks traded in XBT and ETH are much less significant with respect to EUR and USD).     


Summarizing the results obtained with the Granger pairwise analysis, we can say that, on one hand, as one could expect, for a specific time window nodes influence/get influenced each other according to their out/in strength. On the other hand, the structure of the network stays quite stable over consecutive time windows, since the in/out strength values of nodes remain comparable. In this respect, even if in correspondence of important events or shock in the market the structure of the network seems to change, increasing the number of links and their weight, when the effect of those events vanishes the network come back to its original state.

\subsection{O-information}

Let us now describe the results of the analysis of the USD dataset from the point of view of high-order dependencies, using the dynamical O-information introduced in Section \ref{sec:oinformation}. 

\begin{figure}[!ht]
    \centering
\includegraphics[width=0.9\textwidth]{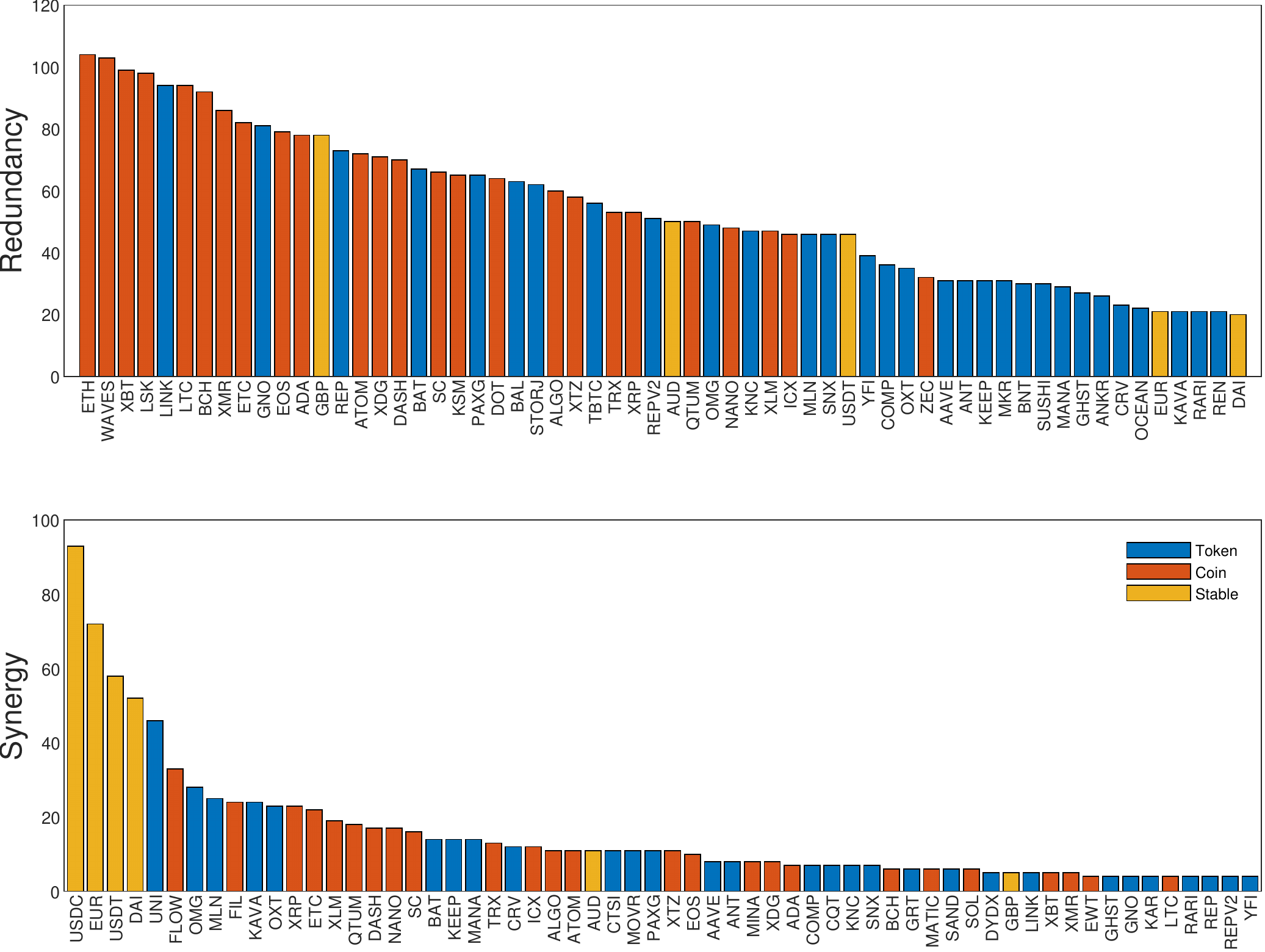}
    \caption{{\bf Most influential nodes for synergy and redundancy in the USD dataset.} The bar plot counts the number of times a cryptocurrency is found in the best multiplets built as explained in the text. Yellow bars indicates stable-coins, red bars coins and blue bars tokens. Only the first $60$ elements are shown.}
    \label{fig:bar_nodes_oinfo}
\end{figure}

A first interesting observation can be made about Figure \ref{fig:bar_nodes_oinfo}, where it is shown the ranking of the presence of each class of cryptocurrency in the best multiplets, both for redundancy and synergy. Here, on the contrary of the pairwise case, one can observe that stable coins are the variables which belong to the highest number of synergistic information circuits, hence they play a major role for high order effects. 

\begin{figure}[!ht]
    \centering
    \includegraphics[width=0.9\textwidth]{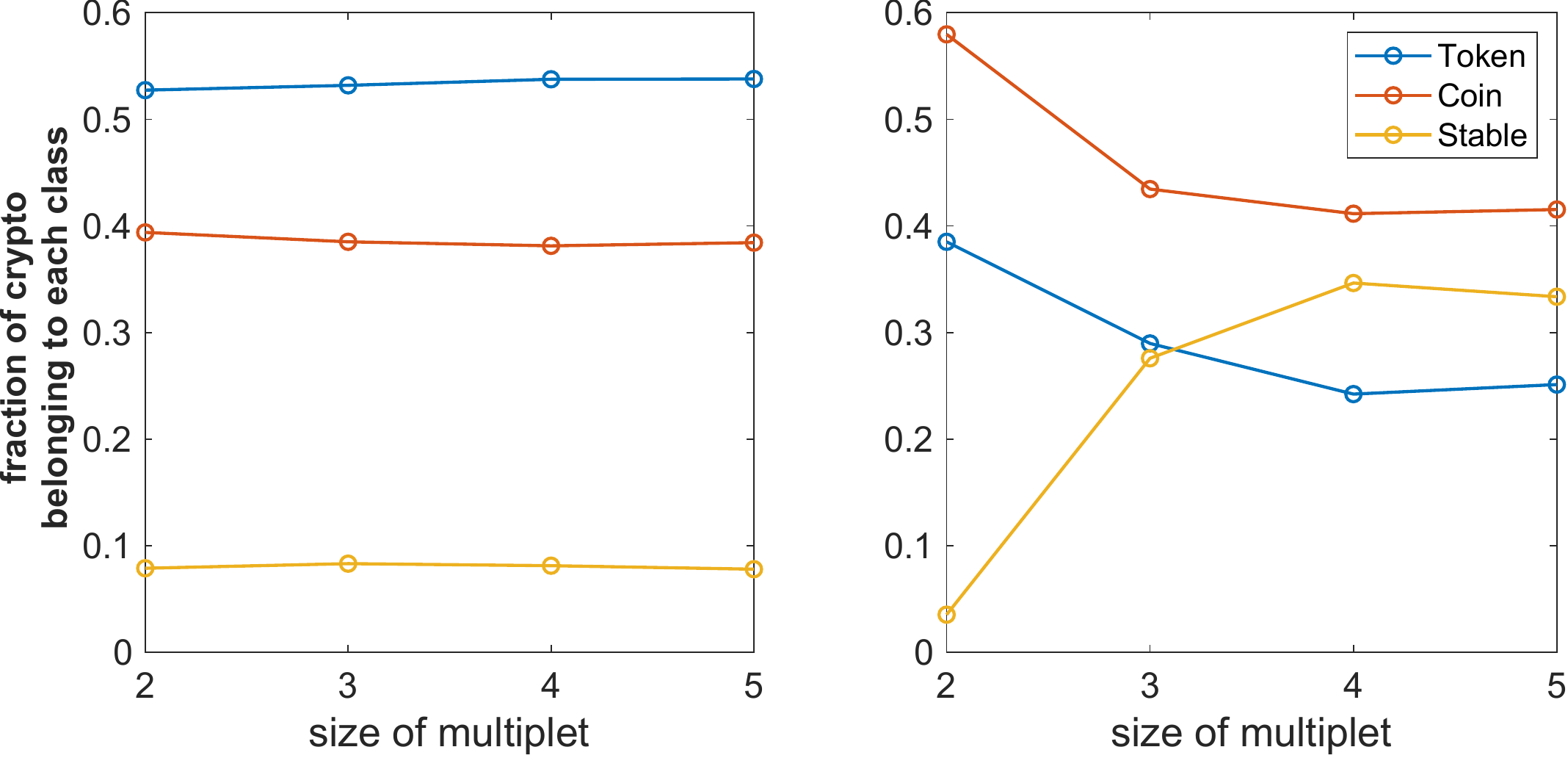}
    \caption{We depict the typical fraction of coin, token and stable-coin in the optimal multiplets $\bm{\tilde{X}}^n$ as a function of their size $n$, for both redundancy (left) and synergy (right). In the redundant case, these weights are similar to the original dataset (composed of 60 tokens, 33 coins and 6 stable-coins), while for synergistic multiplets coins are the most common and the stable-coins are much more present.}
    \label{fig:fraction_multiplets}
\end{figure}

We get other insights by observing the composition of the best multiplets as function of their size. In Figure \ref{fig:fraction_multiplets} we report the typical fraction of elements belonging to the 3 different classes of cryptocurrencies for both redundancy (left panel) and synergy (right panel). Concerning the redundant multiplets, the composition do not depend on the size but only on the class. The highest fraction of elements in these multiplets are tokens, which are also the most common in the original dataset (60/99): high redundancy can be explained from the fact that these cryptocurrency don't have a own blockchain, but are built on top of another crypto-coin. Notice that the lowest fraction are stable coins, which are the least common in the dataset (6/99). 

The situation is very different for the synergistic multiplets (right panel), where the size plays a greater role: actually, the fraction of stable coins increases with the size, while that of the other classes tends to decrease. Interestingly, from size $3$ onwards the best multiplets are populated much more often by coins and, in particular, by the stable coins, even if they are very few in the dataset.

These findings are quite surprising since stable coins have a rather marginal activity in the pairwise Granger causality, as can be seen in Figure \ref{fig:bar_plot_granger}. This result confirms that taking into account high-order dependencies through O-information provides insights that cannot be retrieved with a pairwise analysis.

\subsection{Complementarity of pairwise and high-order descriptions}

In order to better highlight the complementarity between the pairwise and the high-order descriptions, in Figure \ref{fig:comparison_line} we make a comparison between the average values of the pairwise and high-order indicators analyzed so far for the USD dataset. The second and third panel (from the top) address the O-information, showing the values of $d\Omega_{\bm{\tilde{X}^5}\rightarrow y}$ averaged over each target as function of the $104$ weekly windows, for redundancy and synergy respectively. In the bottom panel, we depict the analogous time behavior of Granger causality $\mathcal{F}_{X\rightarrow Y}$ averaged over all the possible pairs. Finally, in the top panel, we show the behavior of total exchanges in US dollars, which can be meant as the total volume of transactions which have been performed during each weekly window.

\begin{figure}[!ht]
    \centering
    \includegraphics[width=0.7\textwidth]{figures/results_all_line.pdf}
    \caption{From top to bottom: the total volume (in US dollar) exchanged in each time window; the average value of redundancy and synergy and the average value of Granger causality $\overline{\mathcal{F}}$ calculated for each time window. The 10-weeks moving average of the signal is represented in red.}
    \label{fig:comparison_line}
\end{figure}

It is quite clear the presence of three successive stages characterized by a different behaviour of the O-information indicators:  (i) the first phase, from January to December 2020, shows high redundancy of the market and low synergy; (ii) the second phase, from January to June 2021, is characterized by a fall of the redundancy and a rise of the synergy; (iii) finally, a third phase can be located between July and December 2021 and is characterized by a stationary dynamics of both synergy and redundancy, which stay relatively high. 
Since synergy may be interpreted as an indicator of the system's complexity, this result suggests that, during the first semester of 2021, the cryptocurrencies network has undergone a transition toward a more complex dynamical landscape. The transition seems to be triggered by the large volume of transactions occurred in the first semester of 2021. 

On the other hand, the behaviour of the Granger causality seems insensitive to such a transition, remaining quite homogeneous during the whole considered period, with the exception of sudden peaks in correspondence of important events - like Covid turbulence in 2020, days of sudden rise in prices etc. Notice, in particular, the peak of 22 Feb 2021, which also appears in the synergy panel and coincides with the first minimum of redundancy. It could be related with the week of anomalous activity already observed in Figure \ref{fig:networks}, maybe the effect of some market movements and announcements of Elon Musk about cryptos and their adoption/dismiss as payment method - as reported by financial press  \footnote{See for example: https://enterprise.press/stories/2021/02/22/what-the-markets-are-doing-on-22-february-2021-32943/}.   

\begin{figure}[!ht]
    \centering
    \includegraphics[width=0.9\textwidth]{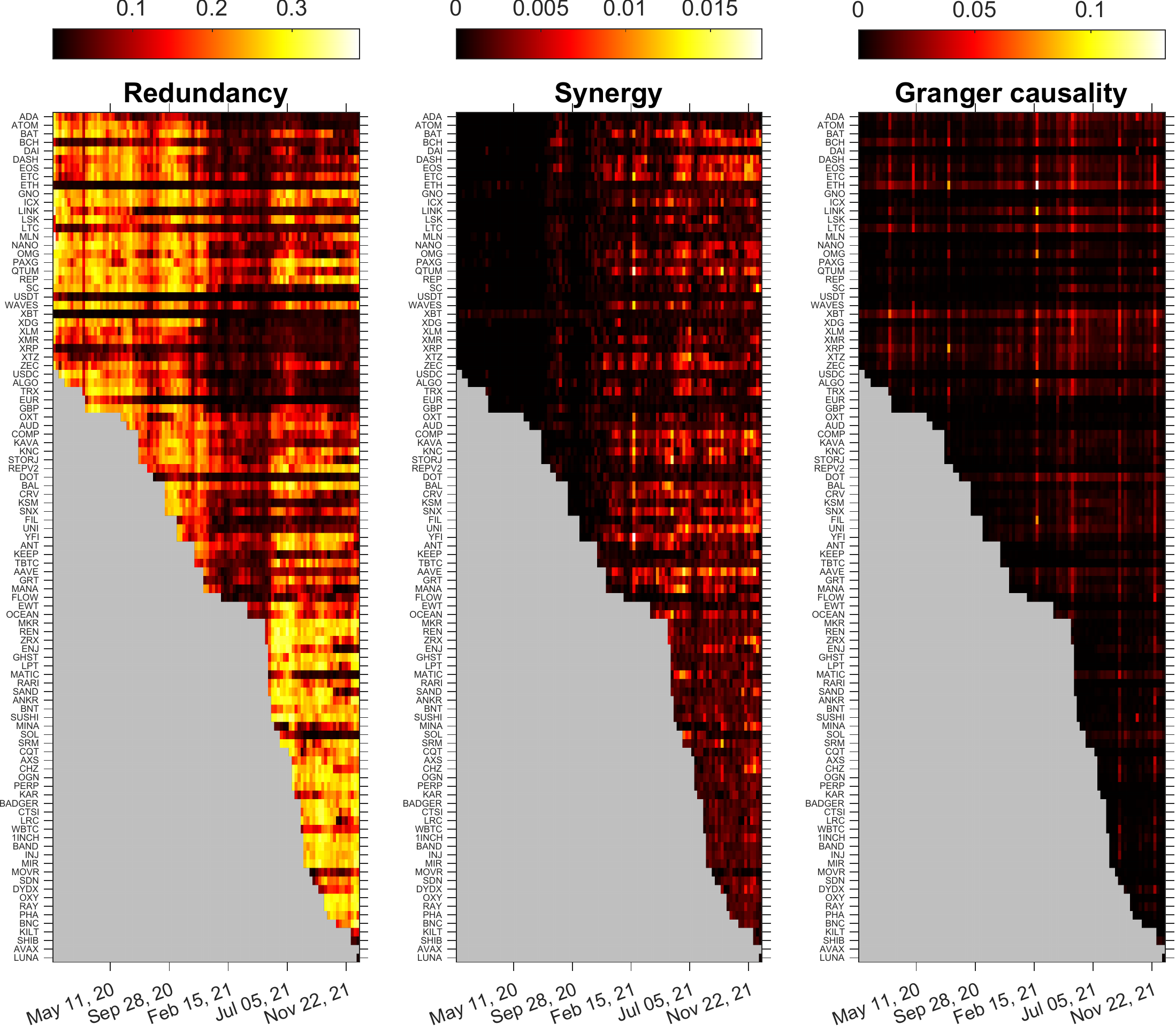}
    \caption{The redundant (left) and synergistic (center) dynamical O-information toward various targets, where each point represents the value of $d\Omega_{{\bm \tilde{X}^5}\rightarrow y}$ of the best multiplet of sources ${\bm \tilde{X}^5}$  that conveys the highest and lowest O-information toward the target $y$. We also depict the out-strength Granger causality (right) for each weekly window. Regions in grey represents windows where that cryptocurrency was not yet listed.}
    \label{fig:comparison_heat}
\end{figure}

A global perspective about the time behaviour of both the O-information and pairwise indicators for each cryptocurrency of the USD dataset can be appreciated in Figure \ref{fig:comparison_heat}, where we report the same quantities of the previous figure but without averaging over all the crypto. In particular, in the left and central panel it is shown the redundancy and synergistic value of dynamical O-information from the best multiplet, respectively. In the right panel, it is shown the Granger causality out strength of each crypto for each weekly time window. Notice that in such a visualization it becomes clear that only about thirty cryptos (the first from the top) are always present in the dataset for the whole considered time period: the others, ordered according to their age, show an increasing number of weekly windows (colored in gray) in which they are not yet listed.    

Focusing on the highest and lowest values of redundancy and synergy, a behaviour analogous to that one observed in Figure \ref{fig:comparison_line} can be easily detected also at this disaggregated level, and the same holds for the Granger causality, thus further confirming the complementarity of the pairwise and the high-order approach.

\section{Conclusions} \label{sec:conclusions}

In this work, we have analysed the logarithmic USD price returns, building a network of traded cryptocurrency by defining the Granger causality between each pairs of cryptocoin in the years 2020 and 2021 as links. We used both pairwise and high order statistical dependencies, as measured by Granger causality and O-information, with a dataset consisting of two classes of currencies (coins and tokens) and six stable coins, for a total of 99 time series. 

We found that, in terms of Granger causality, the network structure is rather stable across time, referring to weekly time windows in the considered period since, according to our analysis, the set of most influential cryptocoins presents small changes over time. The structure of the network may change due to relevant events or shocks within a specific period, but in a short time it tends to restore its original condition.

Turning to consider high-order dependencies, we have analysed how multiplets of cryptocurrencies carry redundant and synergistic information toward the future states of other variables.
In synergistic multiplets, as the size of the multiplets increases, the same does the fraction of stable coin on it, despite their marginal role pointed out by the pairwise analysis.
On the other hand, in the redundant case,  stablecoins fraction is less than 10\% and the multiplets composition does not vary while increasing their size.

We obtained other insights by looking at the temporal trends of the pairwise and high-order indicators. In particular, we compared Granger causality, redundancy and synergy with the time evolution of the total amount of exchanges in US dollars, and we found three successive stages characterized by different behaviour of these indicators.
The first period, between January to December 2020, is characterized by high redundancy and low synergy among cryptocurrencies. Both change directions on the second phase, starting from January to June 2021, where redundancy fall and synergy rise up. At the last phase, located between July and December 2021, both indicators maintains a stationary dynamics.

These results suggest that pairwise and high-order descriptions of complex financial systems provide complementary information, thus, considered together, they represent a very promising tool for the analysis of cryptocurrency trading networks or similar ones.

Conclusively, considering the amount of resources and the relevance of trading activities related to cryptocurrencies, the analysis of their dynamics is a relevant field of investigation aimed at stabilizing their markets. At the end of the day, cryptocurrencies are not financial instruments with an intrinsic value representing productive assets. They have been conceived as mediums of exchange, in markets where their quotation is totally self-referenced and driven by their adoption in transactions (often of unknown type and content). 

Nonetheless, the attracting appeal of positive price jumps and bubbles induces investors to underestimate potential losses. Our analysis of the network of cryptocurrencies might be useful to unveil that while drivers of instability cannot be removed in general, specially in the absence of intrinsic -- fundamental -- values, quotations and prices become extremely risky and unpredictable. This, on a normative side, could inspire policies aimed at fostering stability of markets, anchored more to produced values than to financial speculation.

Matter of factly, some attempts have been done to create a formal market structure for some crypto (i.e. lending platforms such as CoinLoan or Hodlnaut, Binance and Crypto.com as an example of crypto-debit payments card and even Sandbox virual world, where each player can buy objects using cryptocoins and even earning coins by playing the game). The need for a more ``readable'' orientation, aimed at providing priors on the market configuration and on its physiology, shows that despite a financial artefact may reveal appealing in block-chain transactions, it might be perceived as dangerous and too risky.

Our analysis goes to the direction of investigating and understanding what moves the dynamics of such highly volatile crypto currencies, which can be useful from a policy standpoint in order to foster stability and prevent the destruction of relevant financial values.

\appendix

\section{Tables} \label{app:tables}
In Tables \ref{tab:kraken}, \ref{tab:coin_description} and   \ref{tab:data_coverage} we provide more details on the global Kraken dataset used in the analysis.

\begin{table}[b]
	\begin{tabular}{|c| c c |}
		\hline
		{\bf Year} & {\bf number of crypto}  & {\bf number of fiat} \\
		\hline
		2013 & 2 & 2\\
		&(XBT, LTC) & (EUR, USD)\\
		\hline
		2014 & 2 & 4 \\
		& & (+ GBP,JPY only for XBT)\\
		\hline
		2015 & 3 & 5\\
		&(+ ETH) & (+ CAD for XBT and ETH)\\
		\hline
		2016 & 9 & 7\\
		& (+ZEC,XLM,REP,XRP,XDG,ETC)& (+XBT,ETH as fiat) \\
		\hline
		2017 & 16 & 7 \\
		& (+ USDT,DASH,XMR,GNO,MLN,EOS,BCH) & \\
		\hline
		2018 & 19 & 7\\
		&(+ ADA,QTUM,XTZ) & \\
		\hline
		2019 & 30 & 10 \\
		&(...) & (+ USDT,CHF,DAI)\\
		\hline
		2020 & 60 & 11\\
		& (...) & (+AUD)\\
		\hline
		2021 & 99 & 12 \\
& (...)& (+DOT)\\
\hline
\end{tabular}
\caption{Number of cryptocoins and fiat available on Kraken during years. At the beginning of the launch of the Kraken platform, there were available only two cryptos (Bitcoin as XBT and Litecoin as LTC) traded using 2 fiat (Euro and Dollar, using EUR and USD as corresponding tickers). As far as the platform grow up, there were listed more cryptos and fiat, up to 2021 when there are more than 90 cryptos available, and traded using 12 fiat. Since 2016, Bitcoin and Ethereum were also used as ``fiat", to simplify the process of buying and selling.}
\label{tab:kraken}
\end{table}

\begin{table}[]
\fontsize{5}{4}\selectfont
\resizebox{0.95\textwidth}{!}{%
\begin{tabular}{lllll}
\hline

Ticker & Name           & website                 & type & Consensus mechanism    \\ \hline
1INCH                        & 1inch                               & \url{https://1inch.io/}                            & TOKEN                     &                                             \\
AAVE                         & Aave                                & \url{https://aave.com/}                            & TOKEN                     &                                             \\
ADA                          & Cardano                             & \url{https://cardano.org}                          & COIN                      & Proof of Stake                              \\
ALGO                         & Algorand                            & \url{http://algorand.foundation/}                  & COIN                      & Proof of Stake                              \\
ANKR                         & Ankr                                & \url{https://www.ankr.com/}                        & TOKEN                     &                                             \\
ANT                          & Aragon                              & \url{https://aragon.org/}                          & TOKEN                     &                                             \\
ATOM                         & Cosmos                              & \url{https://cosmos.network/}                      & COIN                      & Proof of Stake                              \\
AVAX                         & Avalanche                           & \url{https://avax.network/}                        & COIN                      & Proof of Stake                              \\
AXS                          & Axie Infinity Shards                & \url{https://axieinfinity.com/}                    & TOKEN                     &                                             \\
BADGER                       & Badger DAO                          & \url{https://app.badger.finance/}                  & TOKEN                     &                                             \\
BAL                          & Balancer                            & \url{https://balancer.finance/}                    & TOKEN                     &                                             \\
BAND                         & Band Protocol                       & \url{https://bandprotocol.com/}                    & TOKEN                     &                                             \\
BAT                          & Basic Attention Token               & \url{https://basicattentiontoken.org/}             & TOKEN                     &                                             \\
BCH                          & Bitcoin Cash                        & \url{http://bch.info/}                             & COIN                      & Proof of Work                               \\
BNC                          & Bifrost                             & \url{https://bifrost.finance/}                     & TOKEN                     &                                             \\
BNT                          & Bancor                              & \url{https://bancor.network/}                      & TOKEN                     &                                             \\
CHZ                          & Chiliz                              & \url{https://www.chiliz.com/}                      & TOKEN                     &                                             \\
COMP                         & Compound                            & \url{https://compound.finance/governance/comp}     & TOKEN                     &                                             \\
CQT                          & Covalent                            & \url{https://www.covalenthq.com/}                  & TOKEN                     &                                             \\
CRV                          & Curve                               & \url{https://www.curve.fi/}                        & TOKEN                     &                                             \\
CTSI                         & Cartesi                             & \url{https://cartesi.io/}                          & TOKEN                     &                                             \\
DAI                          & Dai                                 & \url{http://www.makerdao.com/}                     & TOKEN - USD StableCoin    &                                             \\
DASH                         & Dash                                & \url{https://www.dash.org/}                        & COIN                      & Hybrid – PoW \& PoS                         \\
DOT                          & Polkadot                            & \url{https://polkadot.network/}                    & COIN                      & Nominated Proof of Stake                    \\
DYDX                         & dYdX                                & \url{https://dydx.community/}                      & TOKEN                     &                                             \\
ENJ                          & Enjin Coin                          & \url{https://enjin.io/}                            & TOKEN                     &                                             \\
EOS                          & EOSIO                               & \url{https://eos.io}                               & COIN                      & Delegated Proof of Stake                    \\
ETC                          & Ethereum Classic                    & \url{https://ethereumclassic.org/}                 & COIN                      & Proof of Work                               \\
ETH                          & Ethereum                            & \url{https://www.ethereum.org/}                    & COIN                      & Proof of Work                               \\
ETH2.S                       & Ethereum 2.0 Staking                &                                              &                           & Proof of Stake                              \\
EUR                          &                                     &                                              & FIAT                      &                                             \\
EWT                          & Energy Web                          & \url{https://www.energyweb.org/}                   & COIN                      & Proof of Authority                          \\
FIL                          & Filecoin                            & \url{https://filecoin.io/}                         & COIN                      & Proof-of-Replication and Proof-of-Spacetime \\
GBP                          &                                     &                                              & FIAT                      &                                             \\
GHST                         & Aavegotchi                          & \url{https://aavegotchi.com/}                      & TOKEN                     &                                             \\
GNO                          & Gnosis                              & \url{https://gnosis.io/}                         & TOKEN                     &                                             \\
GRT                          & The Graph                           & \url{https://thegraph.com/}                        & TOKEN                     &                                             \\
ICX                          & Icon                                & \url{https://icon.community/}                      & COIN                      & Proof of Stake                              \\
INJ                          & Injective                           & \url{https://injective.com/}                       & TOKEN                     &                                             \\
KAR                          & Karura                              & \url{http://karura.network/}                       & TOKEN                     &                                             \\
KAVA                         & Kava                                & \url{https://www.kava.io/}                         & TOKEN                     &                                             \\
KEEP                         & Keep Network                        & \url{https://keep.network/}                        & TOKEN                     &                                             \\
KILT                         & Kilt Protocol                       & \url{https://kilt.io/}                             & TOKEN                     &                                             \\
KNC                          & Kyber Network & \url{https://kyber.network/} & TOKEN                     &                                             \\
KSM                          & Kusama                              & \url{https://kusama.network/}                      & COIN                      & Nominated Proof of Stake                    \\
LINK                         & Chainlin                            & \url{https://chain.link/}                         & TOKEN                     &                                             \\
LPT                          & Livepeer                            & \url{https://livepeer.org/}                        & TOKEN                     &                                             \\
LRC                          & Loopring                            & \url{https://loopring.org/}                        & TOKEN                     &                                             \\
LSK                          & Lisk                                & \url{https://lisk.com/}                           & COIN                      & Delegated Proof of Stake                    \\
LTC                          & Litecoin                            & \url{https://litecoin.org/}                       & COIN                      & Proof of Work                               \\
LUNA                         & Terra                               & \url{https://terra.money/}                        & COIN                      & Proof of Stake                              \\
MANA                         & Decentraland                        & \url{https://decentraland.org/}                    & TOKEN                     &                                             \\
MATIC                        & Polygon                             & \url{https://polygon.technology/}                  & COIN                      & Proof of Stake                              \\
MINA                         & Mina                                & \url{https://minaprotocol.com/}                    & COIN                      & Zk-SNARK                                    \\
MIR                          & Mirror Protocol                     & \url{https://mirror.finance/}                      & TOKEN                     &                                             \\
MKR                          & Maker                               & \url{https://makerdao.com/}                        & TOKEN                     &                                             \\
MLN                          & Enzyme Finance                      & \url{https://enzyme.finance/}                      & TOKEN                     &                                             \\
MOVR                         & Moonriver                           & \url{https://moonbeam.network/networks/moonriver/} & TOKEN                     &                                             \\
NANO                         & Nano                                & \url{https://nano.org/en}                          & COIN                      & Delegated Proof of Stake                    \\
OCEAN                        & Ocean                               & \url{https://oceanprotocol.com/}                   & TOKEN                     &                                             \\
OGN                          & Origin                              & \url{https://www.originprotocol.com/}              & TOKEN                     &                                             \\
OMG                          & OMG Network                         & \url{https://omg.network/}                         & TOKEN                     &                                             \\
OXT                          & Orchid                              & \url{https://www.orchid.com/}                      & TOKEN                     &                                             \\
OXY                          & Oxygen                              & \url{https://www.oxygen.org/}                      & TOKEN                     &                                             \\
PAXG                         & PAX Gold                            & \url{https://www.paxos.com/paxgold/}               & TOKEN                     &                                             \\
PERP                         & Perpetual Protocol                  & \url{https://perp.com/}                            & TOKEN                     &                                             \\
PHA                          & Phala                               & \url{https://phala.network/}                       & TOKEN                     &                                             \\
QTUM                         & Qtum                                & \url{https://qtum.org/}                            & COIN                      & Mutualized Proof of Stake                   \\
RARI                         & Rarible                             & \url{https://app.rarible.com/rari}                 & TOKEN                     &                                             \\
RAY                          & Raydium                             & \url{https://raydium.io/}                       & TOKEN                     &                                             \\
REN                          & Ren                                 & \url{https://renproject.io/}                       & TOKEN                     &                                             \\
REP                          & Augur                               & \url{http://www.augur.net/}                        & TOKEN                     &                                             \\
REPV2                        & Augur v2                            & \url{http://www.augur.net/}                        & TOKEN                     &                                             \\
SAND                         & Sandbox                             & \url{https://www.sandbox.game/en/}                 & TOKEN                     &                                             \\
SC                           & Siacoin                             & \url{https://sia.tech/}                            & COIN                      & Proof of Work                               \\
SDN                          & Shiden                              & \url{https://shiden.astar.network/}                & TOKEN                     &                                             \\
SHIB                         & Shiba Inu                           & \url{https://shibatoken.com/}                      & TOKEN                     &                                             \\
SNX                          & Synthetix                           & \url{https://www.synthetix.io/}                    & TOKEN                     &                                             \\
SOL                          & Solana                              & \url{https://solana.com/}                          & COIN                      & Proof of Stake                              \\
SRM                          & Serum                               & \url{https://projectserum.com/}                    & TOKEN                     &                                             \\
STORJ                        & Storj                               & \url{https://storj.io/}                            & TOKEN                     &                                             \\
SUSHI                        & Sushi                               & \url{https://sushi.com/}                           & TOKEN                     &                                             \\
TBTC                         & tBTC                                & \url{https://tbtc.network/}                        & TOKEN                     &                                             \\
TRX                          & Tron                                & \url{https://tron.network/}                        & COIN                      & Delegated Proof of Stake                    \\
UNI                          & Uniswap                             & \url{https://uniswap.org/blog/uni/}                & TOKEN                     &                                             \\
USD                          &                                     &                                              & FIAT                      &                                             \\
USDC                         & USD Coin                            & \url{https://www.centre.io/usdc}                   & TOKEN - USD StableCoin    &                                             \\
USDT                         & Tether                              & \url{https://tether.to/}                           & TOKEN - USD StableCoin    &                                             \\
WAVES                        & Waves                               & \url{https://waves.tech/}                          & COIN                      & Leased Proof of Stake                       \\
WBTC                         & Wrapped Bitcoin                     & \url{https://wbtc.network/}                        & TOKEN                     & LpoS                                        \\
XBT                          & Bitcoin                             & \url{https://bitcoin.org/}                         & COIN                      & Proof of Work                               \\
XDG                          & Dogecoin                            & \url{http://dogecoin.com/}                         & COIN                      & Proof of Work                               \\
XLM                          & Stellar                             & \url{https://www.stellar.org/}                     & COIN                      & Stellar Consensus Protocol                  \\
XMR                          & Monero                              & \url{https://www.getmonero.org/}                   & COIN                      & Proof of Work                               \\
XRP                          & Ripple                              & \url{https://ripple.com/}                          & COIN                      & Ripple consensus                            \\
XTZ                          & Tezos                               & \url{https://www.tezos.com/}                       & COIN                      & Liquid Proof of Stake                       \\
YFI                          & yEarn                               & \url{https://yearn.finance/}                       & TOKEN                     &                                             \\
ZEC                          & Zcash                               & \url{https://z.cash/}                              & COIN                      & Zk-SNARK                                    \\
ZRX                          & 0x                                  & \url{https://0x.org/}                              & TOKEN                     &                                           
\end{tabular}
}

\caption{Tickers description. Each row provides some information on the ticker, such as the name of the coin, the website of the project, its type (COIN for cryptocoin which hold their own blockchain, TOKEN for cryptos built on top of another cryptocoin, fiat for currencies and TOKEN - USD Stablecoin for token hooked to the USD dollar price) and what algorithm it is used as consensus mechanism (for COINS only).
}
\label{tab:coin_description}
\end{table}

\begin{table}[]
    \centering
    \resizebox{\textwidth}{!}{%
    \begin{tabular}{|l|c|c|c|c|c|c|c|c|c|c|c|c|c|c|c|c|c|c|c|c|c|c|c|c|c|c|c|}
    \hline
     & & \multicolumn{13}{c|}{\centering 2020} & \multicolumn{13}{c|}{\centering 2021} \\ \hline
crypto & firstday & EUR & USD & GBP & CAD & AUD & CHF & DAI & JPY & USDT & USDC & XBT & ETH & DOT & EUR & USD & GBP & CAD & AUD & CHF & DAI & JPY & USDT & USDC & XBT & ETH & DOT \\ 
\hline
1INCH & 20210810 &  &  &  &  &  &  &  &  &  &  &  &  &  & 144 & 144 &  &  &  &  &  &  &  &  &  &  & \\ 
AAVE & 20201215 & 17 & 17 & 17 &  &  &  &  &  &  &  & 17 & 17 &  & 365 & 365 & 365 &  & 345 &  &  &  &  &  & 365 & 365 & \\ 
ADA & 20191230 & 368 & 368 &  &  &  &  &  &  & 78 &  & 368 & 368 &  & 365 & 365 & 345 &  & 345 &  &  &  & 365 &  & 365 & 365 & \\ 
ALGO & 20200122 & 345 & 345 &  &  &  &  &  &  &  &  & 346 & 345 &  & 365 & 365 & 345 &  &  &  &  &  &  &  & 365 & 365 & \\ 
ANKR & 20210524 &  &  &  &  &  &  &  &  &  &  &  &  &  & 222 & 222 & 222 &  &  &  &  &  &  &  & 222 &  & \\ 
ANT & 20201124 & 38 & 38 &  &  &  &  &  &  &  &  & 38 & 38 &  & 365 & 365 &  &  &  &  &  &  &  &  & 365 & 365 & \\ 
ATOM & 20191230 & 368 & 368 &  &  &  &  &  &  &  &  & 368 & 368 &  & 365 & 365 & 345 &  & 345 &  &  &  &  &  & 365 & 365 & \\ 
AUD & 20200616 &  & 199 &  &  &  &  &  & 199 &  &  &  &  &  &  & 365 &  &  &  &  &  & 363 &  &  &  &  & \\ 
AVAX & 20211221 &  &  &  &  &  &  &  &  &  &  &  &  &  & 11 & 11 &  &  &  &  &  &  &  &  &  &  & \\ 
AXS & 20210713 &  &  &  &  &  &  &  &  &  &  &  &  &  & 172 & 172 &  &  &  &  &  &  &  &  &  &  & \\ 
BADGER & 20210803 &  &  &  &  &  &  &  &  &  &  &  &  &  & 151 & 151 &  &  &  &  &  &  &  &  &  &  & \\ 
BAL & 20200917 & 106 & 106 &  &  &  &  &  &  &  &  & 106 & 106 &  & 365 & 365 &  &  &  &  &  &  &  &  & 365 & 365 & \\ 
BAND & 20210810 &  &  &  &  &  &  &  &  &  &  &  &  &  & 144 & 144 &  &  &  &  &  &  &  &  &  &  & \\ 
BAT & 20191230 & 368 & 368 &  &  &  &  &  &  &  &  & 368 & 368 &  & 365 & 365 &  &  &  &  &  &  &  &  & 365 & 365 & \\ 
BCH & 20200616 & 368 & 368 & 246 &  & 199 &  &  & 71 & 246 &  & 368 & 246 &  & 365 & 365 & 365 &  & 365 &  &  & 365 & 365 &  & 365 & 365 & \\ 
BNC & 20211026 &  &  &  &  &  &  &  &  &  &  &  &  &  & 67 & 67 &  &  &  &  &  &  &  &  &  &  & \\ 
BNT & 20210524 &  &  &  &  &  &  &  &  &  &  &  &  &  & 222 & 222 & 222 &  &  &  &  &  &  &  & 222 &  & \\ 
CHZ & 20210713 &  &  &  &  &  &  &  &  &  &  &  &  &  & 172 & 172 &  &  &  &  &  &  &  &  &  &  & \\ 
COMP & 20200715 & 170 & 170 &  &  &  &  &  &  &  &  & 170 & 170 &  & 365 & 365 &  &  &  &  &  &  &  &  & 365 & 365 & \\ 
CQT & 20210706 &  &  &  &  &  &  &  &  &  &  &  &  &  & 179 & 179 &  &  &  &  &  &  &  &  &  &  & \\ 
CRV & 20200917 & 106 & 106 &  &  &  &  &  &  &  &  & 106 & 106 &  & 365 & 365 &  &  &  &  &  &  &  &  & 365 & 365 & \\ 
CTSI & 20210803 &  &  &  &  &  &  &  &  &  &  &  &  &  & 151 & 151 &  &  &  &  &  &  &  &  &  &  & \\ 
DAI & 20191230 & 368 & 368 &  &  &  &  &  &  & 368 &  &  &  &  & 365 & 365 &  &  &  &  &  &  & 365 &  &  &  & \\ 
DASH & 20191230 & 368 & 368 &  &  &  &  &  &  &  &  & 368 &  &  & 365 & 365 &  &  &  &  &  &  &  &  & 365 &  & \\ 
DOT & 20201015 & 136 & 136 &  &  &  &  &  &  & 78 &  & 136 & 136 &  & 365 & 365 & 345 &  & 345 &  &  &  & 365 &  & 365 & 365 & \\ 
DYDX & 20210914 &  &  &  &  &  &  &  &  &  &  &  &  &  & 109 & 109 &  &  &  &  &  &  &  &  &  &  & \\ 
ENJ & 20210520 &  &  &  &  &  &  &  &  &  &  &  &  &  & 226 & 226 & 225 &  &  &  &  &  &  &  & 226 &  & \\ 
EOS & 20201015 & 368 & 368 &  &  &  &  &  &  & 78 &  & 368 & 368 &  & 365 & 365 &  &  &  &  &  &  & 365 &  & 365 & 365 & \\ 
ETC & 20191230 & 368 & 368 &  &  &  &  &  &  &  &  & 368 & 368 &  & 365 & 365 &  &  &  &  &  &  &  &  & 365 & 365 & \\ 
ETH & 20191230 & 368 & 368 & 368 & 368 & 198 & 368 & 368 & 368 & 368 & 359 & 368 &  &  & 365 & 365 & 365 & 365 & 365 & 365 & 365 & 365 & 365 & 365 & 365 &  & \\ 
ETH2.S & 20211001 &  &  &  &  &  &  &  &  &  &  &  &  &  &  &  &  &  &  &  &  &  &  &  &  & 92 & \\ 
EUR & 20200617 &  & 295 & 295 & 295 & 198 & 295 &  & 295 &  &  &  &  &  &  & 365 & 365 & 365 & 365 & 365 &  & 365 &  &  &  &  & \\ 
EWT & 20210401 &  &  &  &  &  &  &  &  &  &  &  &  &  & 275 & 275 & 275 &  &  &  &  &  &  &  & 275 &  & \\ 
FIL & 20201015 & 78 & 78 &  &  &  &  &  &  &  &  & 78 & 78 &  & 365 & 365 & 345 &  & 345 &  &  &  &  &  & 365 & 365 & \\ 
FLOW & 20210127 &  &  &  &  &  &  &  &  &  &  &  &  &  & 339 & 339 & 339 &  &  &  &  &  &  &  & 339 & 339 & \\ 
GBP & 20200312 &  & 295 &  &  &  &  &  &  &  &  &  &  &  &  & 365 &  &  &  &  &  &  &  &  &  &  & \\ 
GHST & 20210517 &  &  &  &  &  &  &  &  &  &  &  &  &  & 229 & 229 & 229 &  &  &  &  &  &  &  & 229 &  & \\ 
GNO & 20191230 & 368 & 368 &  &  &  &  &  &  &  &  & 368 & 368 &  & 365 & 365 &  &  &  &  &  &  &  &  & 365 & 365 & \\ 
GRT & 20201218 & 14 & 14 &  &  &  &  &  &  &  &  & 14 & 14 &  & 365 & 365 & 345 &  & 345 &  &  &  &  &  & 365 & 365 & \\ 
ICX & 20191230 & 368 & 368 &  &  &  &  &  &  &  &  & 368 & 368 &  & 365 & 365 &  &  &  &  &  &  &  &  & 365 & 365 & \\ 
INJ & 20210810 &  &  &  &  &  &  &  &  &  &  &  &  &  & 144 & 144 &  &  &  &  &  &  &  &  &  &  & \\ 
KAR & 20210720 &  &  &  &  &  &  &  &  &  &  &  &  &  & 165 & 165 &  &  &  &  &  &  &  &  &  &  & \\ 
KAVA & 20200715 & 170 & 170 &  &  &  &  &  &  &  &  & 170 & 170 &  & 365 & 365 &  &  &  &  &  &  &  &  & 365 & 365 & \\ 
KEEP & 20201124 & 38 & 38 &  &  &  &  &  &  &  &  & 38 & 38 &  & 365 & 365 &  &  &  &  &  &  &  &  & 365 & 365 & \\ 
KILT & 20211130 &  &  &  &  &  &  &  &  &  &  &  &  &  & 32 & 32 &  &  &  &  &  &  &  &  &  &  & \\ 
KNC & 20200715 & 170 & 170 &  &  &  &  &  &  &  &  & 170 & 170 &  & 365 & 365 &  &  &  &  &  &  &  &  & 365 & 365 & \\ 
KSM & 20210121 & 106 & 106 &  &  &  &  &  &  &  &  & 106 & 106 &  & 365 & 365 & 345 &  & 344 &  &  &  &  &  & 365 & 365 & 267\\ 
LINK & 20210121 & 368 & 368 &  &  &  &  &  &  & 78 &  & 368 & 368 &  & 365 & 365 & 345 &  & 345 &  &  &  & 365 &  & 365 & 365 & \\ 
LPT & 20210520 &  &  &  &  &  &  &  &  &  &  &  &  &  & 226 & 226 & 226 &  &  &  &  &  &  &  & 226 &  & \\ 
LRC & 20210803 &  &  &  &  &  &  &  &  &  &  &  &  &  & 151 & 151 &  &  &  &  &  &  &  &  &  &  & \\ 
LSK & 20191230 & 368 & 368 &  &  &  &  &  &  &  &  & 368 & 368 &  & 365 & 365 &  &  &  &  &  &  &  &  & 365 & 365 & \\ 
LTC & 20191230 & 368 & 368 & 245 &  & 198 &  &  & 71 & 246 &  & 368 & 246 &  & 365 & 365 & 365 &  & 365 &  &  & 365 & 365 &  & 365 & 365 & \\ 
LUNA & 20211216 &  &  &  &  &  &  &  &  &  &  &  &  &  & 16 & 16 &  &  &  &  &  &  &  &  &  &  & \\ 
MANA & 20201215 & 17 & 17 &  &  &  &  &  &  &  &  & 17 & 17 &  & 365 & 365 &  &  &  &  &  &  &  &  & 365 & 365 & \\ 
MATIC & 20210517 &  &  &  &  &  &  &  &  &  &  &  &  &  & 229 & 229 & 229 &  &  &  &  &  &  &  & 229 &  & \\ 
MINA & 20210601 &  &  &  &  &  &  &  &  &  &  &  &  &  & 214 & 214 & 214 &  &  &  &  &  &  &  & 214 &  & \\ 
MIR & 20210810 &  &  &  &  &  &  &  &  &  &  &  &  &  & 144 & 144 &  &  &  &  &  &  &  &  &  &  & \\ 
MKR & 20210514 &  &  &  &  &  &  &  &  &  &  &  &  &  & 232 & 232 & 232 &  &  &  &  &  &  &  & 232 &  & \\ 
MLN & 20191230 & 368 & 368 &  &  &  &  &  &  &  &  & 368 & 368 &  & 365 & 365 &  &  &  &  &  &  &  &  & 365 & 365 & \\ 
MOVR & 20210827 &  &  &  &  &  &  &  &  &  &  &  &  &  & 127 & 127 &  &  &  &  &  &  &  &  &  &  & \\ 
NANO & 20191230 & 368 & 368 &  &  &  &  &  &  &  &  & 368 & 368 &  & 365 & 365 &  &  &  &  &  &  &  &  & 365 & 365 & \\ 
OCEAN & 20210401 &  &  &  &  &  &  &  &  &  &  &  &  &  & 275 & 275 & 275 &  &  &  &  &  &  &  & 275 &  & \\ 
OGN & 20210713 &  &  &  &  &  &  &  &  &  &  &  &  &  & 172 & 172 &  &  &  &  &  &  &  &  &  &  & \\ 
OMG & 20191230 & 368 & 368 &  &  &  &  &  &  &  &  & 368 & 368 &  & 365 & 365 &  &  &  &  &  &  &  &  & 365 & 365 & \\ 
OXT & 20200602 & 213 & 213 &  &  &  &  &  &  &  &  & 213 & 213 &  & 365 & 365 &  &  &  &  &  &  &  &  & 365 & 365 & \\ 
OXY & 20210928 &  &  &  &  &  &  &  &  &  &  &  &  &  & 95 & 95 &  &  &  &  &  &  &  &  &  &  & \\ 
PAXG & 20191230 & 368 & 368 &  &  &  &  &  &  &  &  & 368 & 368 &  & 365 & 365 &  &  &  &  &  &  &  &  & 365 & 365 & \\ 
PERP & 20210713 &  &  &  &  &  &  &  &  &  &  &  &  &  & 172 & 172 &  &  &  &  &  &  &  &  &  &  & \\ 
PHA & 20211008 &  &  &  &  &  &  &  &  &  &  &  &  &  & 85 & 85 &  &  &  &  &  &  &  &  &  &  & \\ 
QTUM & 20191230 & 368 & 368 &  &  &  &  &  &  &  &  & 368 & 368 &  & 365 & 365 &  &  &  &  &  &  &  &  & 365 & 365 & \\ 
RARI & 20210517 &  &  &  &  &  &  &  &  &  &  &  &  &  & 229 & 229 & 229 &  &  &  &  &  &  &  & 229 &  & \\ 
RAY & 20210928 &  &  &  &  &  &  &  &  &  &  &  &  &  & 95 & 95 &  &  &  &  &  &  &  &  &  &  & \\ 
REN & 20210514 &  &  &  &  &  &  &  &  &  &  &  &  &  & 232 & 232 & 232 &  &  &  &  &  &  &  & 232 &  & \\ 
REP & 20191230 & 368 & 368 &  &  &  &  &  &  &  &  & 368 & 368 &  & 365 & 365 &  &  &  &  &  &  &  &  & 365 & 365 & \\ 
REPV2 & 20200804 & 150 & 150 &  &  &  &  &  &  &  &  & 150 & 150 &  & 365 & 365 &  &  &  &  &  &  &  &  & 365 & 365 & \\ 
SAND & 20210520 &  &  &  &  &  &  &  &  &  &  &  &  &  & 226 & 226 & 226 &  &  &  &  &  &  &  & 226 &  & \\ 
SC & 20191230 & 368 & 368 &  &  &  &  &  &  &  &  & 368 & 368 &  & 365 & 365 &  &  &  &  &  &  &  &  & 365 & 365 & \\ 
SDN & 20210902 &  &  &  &  &  &  &  &  &  &  &  &  &  & 121 & 121 &  &  &  &  &  &  &  &  &  &  & \\ 
SHIB & 20211130 &  &  &  &  &  &  &  &  &  &  &  &  &  & 32 & 32 &  &  &  &  &  &  &  &  &  &  & \\ 
SNX & 20200917 & 106 & 106 &  &  &  &  &  &  &  &  & 106 & 106 &  & 365 & 365 & 345 &  & 345 &  &  &  &  &  & 365 & 365 & \\ 
SOL & 20210617 &  &  &  &  &  &  &  &  &  &  &  &  &  & 198 & 198 & 198 &  &  &  &  &  &  &  & 198 &  & \\ 
SRM & 20210617 &  &  &  &  &  &  &  &  &  &  &  &  &  & 198 & 198 & 198 &  &  &  &  &  &  &  & 198 &  & \\ 
STORJ & 20200715 & 170 & 170 &  &  &  &  &  &  &  &  & 170 & 170 &  & 365 & 365 &  &  &  &  &  &  &  &  & 365 & 365 & \\ 
SUSHI & 20210524 &  &  &  &  &  &  &  &  &  &  &  &  &  & 222 & 222 & 222 &  &  &  &  &  &  &  & 222 &  & \\ 
TBTC & 20201125 & 38 & 37 &  &  &  &  &  &  &  &  & 37 & 38 &  & 365 & 365 &  &  &  &  &  &  &  &  & 365 & 365 & \\ 
TRX & 20200305 & 302 & 302 &  &  &  &  &  &  &  &  & 302 & 302 &  & 365 & 365 &  &  &  &  &  &  &  &  & 365 & 365 & \\ 
UNI & 20201015 & 78 & 78 &  &  &  &  &  &  &  &  & 78 & 78 &  & 365 & 365 &  &  &  &  &  &  &  &  & 365 & 365 & \\ 
USD & 20200312 &  &  &  & 295 &  & 295 &  & 295 &  &  &  &  &  &  &  &  & 365 &  & 365 &  & 365 &  &  &  &  & \\ 
USDC & 20200108 & 359 & 359 &  &  &  &  &  &  & 359 &  &  &  &  & 365 & 365 & 345 &  & 345 &  &  &  & 365 &  &  &  & \\ 
USDT & 20191230 & 368 & 368 & 368 & 368 & 198 & 245 &  & 244 &  &  &  &  &  & 365 & 365 & 365 & 365 & 365 & 365 &  & 365 &  &  &  &  & \\ 
WAVES & 20191230 & 368 & 368 &  &  &  &  &  &  &  &  & 368 & 368 &  & 365 & 365 &  &  &  &  &  &  &  &  & 365 & 365 & \\ 
WBTC & 20210825 &  &  &  &  &  &  &  &  &  &  &  &  &  & 151 & 151 &  &  &  &  &  &  &  &  & 129 &  & \\ 
XBT & 20191230 & 368 & 368 & 368 & 368 & 198 & 368 & 368 & 368 & 368 & 359 &  &  &  & 365 & 365 & 365 & 365 & 365 & 365 & 365 & 365 & 365 & 365 &  &  & \\ 
XDG & 20191230 & 368 & 368 &  &  &  &  &  &  &  &  & 368 &  &  & 365 & 365 &  &  &  &  &  &  & 196 &  & 365 &  & \\ 
XLM & 20191230 & 368 & 368 &  &  &  &  &  &  &  &  & 368 &  &  & 365 & 365 & 345 &  & 345 &  &  &  &  &  & 365 &  & \\ 
XMR & 20191230 & 368 & 368 &  &  &  &  &  &  &  &  & 368 &  &  & 365 & 365 &  &  &  &  &  &  &  &  & 365 &  & \\ 
XRP & 20191230 & 368 & 368 & 246 & 368 & 198 &  &  & 368 & 246 &  & 368 & 246 &  & 365 & 365 & 365 & 365 & 365 &  &  & 365 & 365 &  & 365 & 365 & \\ 
XTZ & 20191230 & 368 & 368 &  &  &  &  &  &  &  &  & 368 & 368 &  & 365 & 365 & 345 &  & 345 &  &  &  &  &  & 365 & 365 & \\ 
YFI & 20201015 & 78 & 78 &  &  &  &  &  &  &  &  & 78 & 78 &  & 365 & 365 & 345 &  & 345 &  &  &  &  &  & 365 & 365 & \\ 
ZEC & 20191230 & 368 & 368 &  &  &  &  &  &  &  &  & 368 &  &  & 365 & 365 &  &  &  &  &  &  &  &  & 365 &  & \\ 
ZRX & 20210514 &  &  &  &  &  &  &  &  &  &  &  &  &  & 232 & 232 & 232 &  &  &  &  &  &  &  & 232 &  & \\ 
\hline

    \end{tabular}}
    \caption{Data coverage on 2020 (starting on Monday 30/12/2019) and 2021. Rows are cryptocoins, columns correspond to fiat coin used to trade. The crypto column reports the Ticker used to list the cryptocoin on the Kraken Exchange. Firstday column show when the cryptopair appear for the first time (on the 2020 or 2021) using the format yyymmdd. For each fiat, are shown the number of days available for the crypto-fiat pair in 2020 and in 2021.}
    \label{tab:data_coverage}
\end{table}

\section{Results for other fiat} \label{app:otherfiat}
\begin{figure}[!ht]
\centering
\begin{tabular}{cc}
\subcaptionbox{USD networks, 99 nodes \label{fig:corr_USD}}{\includegraphics[width =0.5\textwidth]{figures/2020_2021_Correlation_between_GC_networks_USD.pdf}} & \subcaptionbox{EUR networks, 96 nodes \label{fig:corr_EUR}}{\includegraphics[width = 0.5\textwidth]{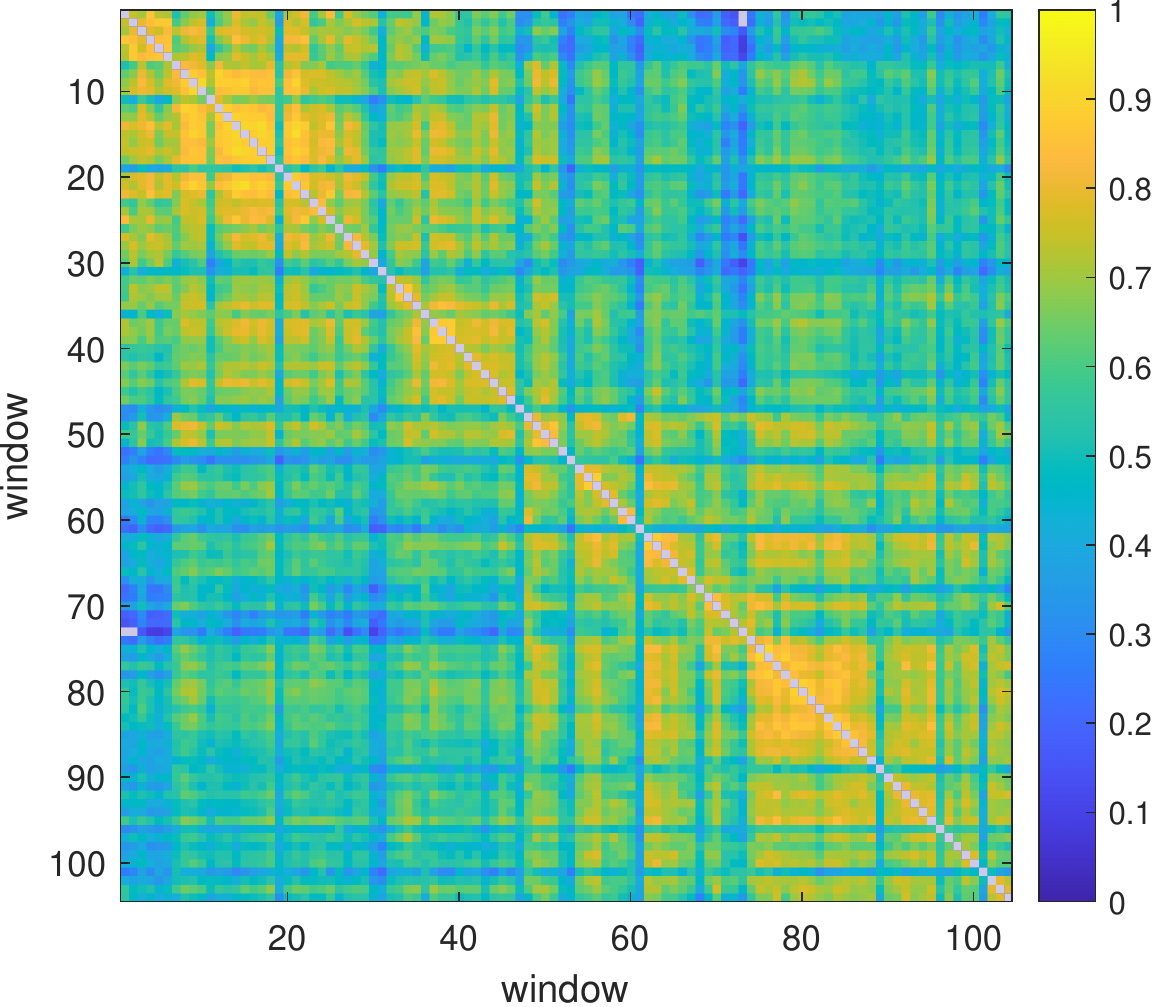}}\\\\
\subcaptionbox{XBT networks, 68 nodes \label{fig:corr_XBT}}{\includegraphics[width =0.5\textwidth]{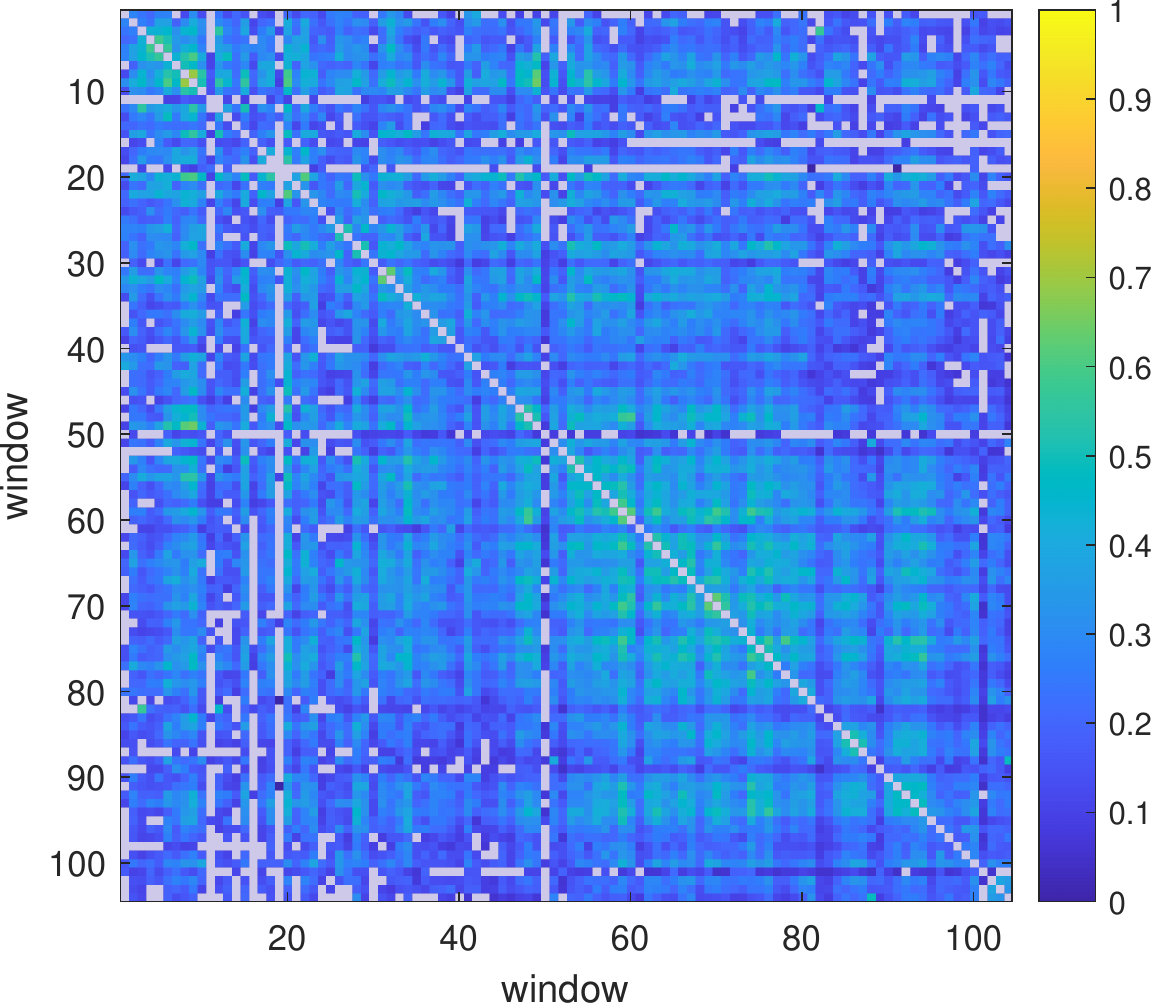}} & \subcaptionbox{ETH networks, 44 nodes \label{fig:corr_ETH}}{\includegraphics[width = 0.5\textwidth]{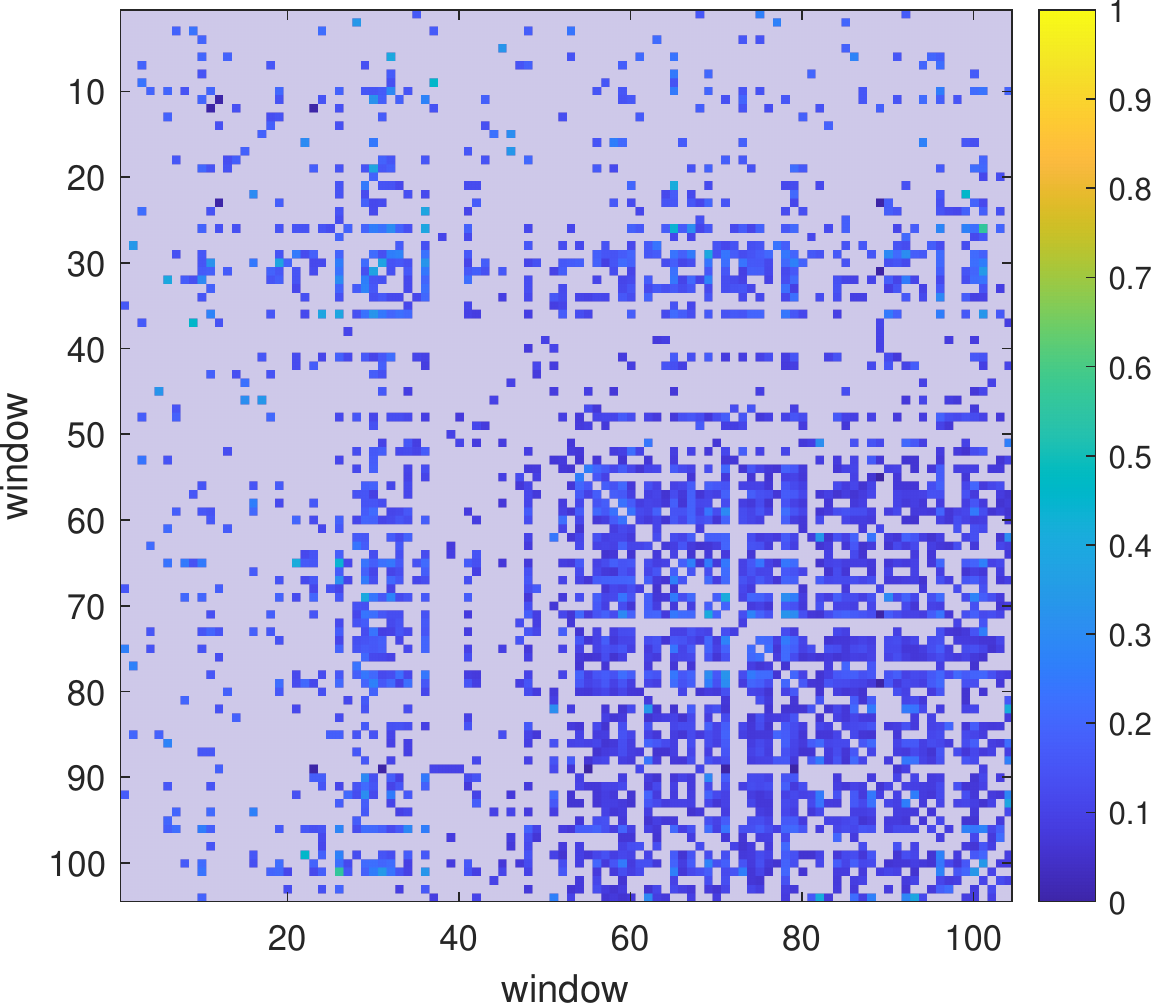}}
\end{tabular}
\caption{
{\bf The Pearson correlation between adjacency matrices at different time windows.}
The element $c_{ij} = \rho(A^i,A^j)$  of each of these matrices is the Pearson correlation between the vectorized adjacency matrices at windows $i$ and $j$. High values of $c_{ij}$ indicate that the structure of the network at windows $i$ and $j$ is very similar.
\ref{fig:corr_USD} and \ref{fig:corr_EUR} are the correlation between cryptos networks traded using respectively USD and EUR. As it is possible to see they are quite similar, except for some window where correlation is discontinued on USD. 
Panels \ref{fig:corr_XBT} and \ref{fig:corr_ETH} show the correlation between networks using XBT and ETH as fiat. Despite the number of nodes being similar to USD and EUR networks, the correlation profile is quite low compared to them. Correlations that are not significant ($p > 0.01$), are indicated in gray.
}\label{fig:corr_2020_2021_all}
\end{figure}

For completeness, in Figure \ref{fig:corr_2020_2021_all} we show the time correlation matrices obtained by using also the other fiat currencies of the dataset.
Concerning the stability of the network, we see that results for EUR \ref{fig:corr_EUR} is quite similar to USD \ref{fig:corr_USD}. On the other hand, in Figures \ref{fig:corr_XBT} and \ref{fig:corr_ETH} it can be seen that the matrix of correlation for XBT and ETH has low and often non-significant values, indicating that the network is quite unstable.

\clearpage
\bibliographystyle{unsrt}
\bibliography{biblio}

\begin{thebibliography}{10}

\bibitem{yahoofinance1}
Yahoo finance bitcoin data in usd for 2017.
\newblock \url{https://yhoo.it/3xJ2QR2} , accessed on $13^{th} $ June, 2022.

\bibitem{yahoofinance2}
Yahoo finance bitcoin data in usd, first quarter of 2018.
\newblock \url{https://yhoo.it/3xKBV7C}, accessed on $13^{th} $ June, 2022.

\bibitem{pearl_causality_2009}
Judea Pearl.
\newblock {\em Causality}.
\newblock Cambridge university press, 2009.

\bibitem{wiener_modern_1956}
Norbert Wiener and Edwin Beckenbach.
\newblock {\em Modern mathematics for engineers}.
\newblock 1956.

\bibitem{granger_economic_1963}
Clive William~John Granger.
\newblock Economic processes involving feedback.
\newblock {\em Information and control}, 6(1):28--48, 1963.
\newblock Publisher: Elsevier.

\bibitem{granger_investigating_1969}
Clive~WJ Granger.
\newblock Investigating causal relations by econometric models and
  cross-spectral methods.
\newblock {\em Econometrica: journal of the Econometric Society}, pages
  424--438, 1969.
\newblock Publisher: JSTOR.

\bibitem{stokes2017study}
Patrick~A Stokes and Patrick~L Purdon.
\newblock A study of problems encountered in granger causality analysis from a
  neuroscience perspective.
\newblock {\em Proceedings of the national academy of sciences},
  114(34):E7063--E7072, 2017.

\bibitem{kodra2011exploring}
Evan Kodra, Snigdhansu Chatterjee, and Auroop~R Ganguly.
\newblock Exploring granger causality between global average observed time
  series of carbon dioxide and temperature.
\newblock {\em Theoretical and applied climatology}, 104(3):325--335, 2011.

\bibitem{mediano2022integrated}
Pedro~AM Mediano, Fernando~E Rosas, Juan~Carlos Farah, Murray Shanahan, Daniel
  Bor, and Adam~B Barrett.
\newblock Integrated information as a common signature of dynamical and
  information-processing complexity.
\newblock {\em Chaos: An Interdisciplinary Journal of Nonlinear Science},
  32(1):013115, 2022.

\bibitem{scagliarini2020synergistic}
Tomas Scagliarini, Luca Faes, Daniele Marinazzo, Sebastiano Stramaglia, and
  Rosario~N Mantegna.
\newblock Synergistic information transfer in the global system of financial
  markets.
\newblock {\em Entropy}, 22(9):1000, 2020.

\bibitem{williams2010nonnegative}
Paul~L Williams and Randall~D Beer.
\newblock Nonnegative decomposition of multivariate information.
\newblock {\em arXiv preprint arXiv:1004.2515}, 2010.

\bibitem{rosas2019quantifying}
Fernando~E Rosas, Pedro~AM Mediano, Michael Gastpar, and Henrik~J Jensen.
\newblock Quantifying high-order interdependencies via multivariate extensions
  of the mutual information.
\newblock {\em Physical Review E}, 100(3):032305, 2019.

\bibitem{luppi2022}
Andrea~I. Luppi, Pedro A.~M. Mediano, Fernando~E. Rosas, Negin Holland, Tim~D.
  Fryer, John~T. O’Brien, James~B. Rowe, David~K. Menon, Daniel Bor, and
  Emmanuel~A. Stamatakis.
\newblock {A synergistic core for human brain evolution and cognition}.
\newblock {\em Nature Neuroscience}, 4(3):910--924, 2022.

\bibitem{nuzzi2020}
Davide Nuzzi, Mario Pellicoro, Leonardo Angelini, Daniele Marinazzo, and
  Sebastiano Stramaglia.
\newblock {Synergistic information in a dynamical model implemented on the
  human structural connectome reveals spatially distinct associations with
  age}.
\newblock {\em Network Neuroscience}, 4(3):910--924, 2020.

\bibitem{aslanidis2019analysis}
Nektarios Aslanidis, Aurelio~F Bariviera, and Oscar Mart{\'\i}nez-Iba{\~n}ez.
\newblock An analysis of cryptocurrencies conditional cross correlations.
\newblock {\em Finance Research Letters}, 31:130--137, 2019.

\bibitem{kruckeberg2019cryptocurrencies}
Sinan Kr{\"u}ckeberg and Peter Scholz.
\newblock Cryptocurrencies as an asset class.
\newblock In {\em Cryptofinance and mechanisms of exchange}, pages 1--28.
  Springer, 2019.

\bibitem{briere2015virtual}
Marie Briere, Kim Oosterlinck, and Ariane Szafarz.
\newblock Virtual currency, tangible return: Portfolio diversification with
  bitcoin.
\newblock {\em Journal of Asset Management}, 16(6):365--373, 2015.

\bibitem{elendner2018cross}
Hermann Elendner, Simon Trimborn, Bobby Ong, and Teik~Ming Lee.
\newblock The cross-section of crypto-currencies as financial assets: Investing
  in crypto-currencies beyond bitcoin.
\newblock In {\em Handbook of Blockchain, Digital Finance, and Inclusion,
  Volume 1}, pages 145--173. Elsevier, 2018.

\bibitem{koutmos_return_2018}
Dimitrios Koutmos.
\newblock Return and volatility spillovers among cryptocurrencies.
\newblock {\em Economics Letters}, 173:122--127, 2018.
\newblock Publisher: Elsevier.

\bibitem{luu_duc_huynh_spillover_2019}
Toan Luu Duc~Huynh.
\newblock Spillover risks on cryptocurrency markets: {A} look from {VAR}-{SVAR}
  granger causality and student’st copulas.
\newblock {\em Journal of Risk and Financial Management}, 12(2):52, 2019.
\newblock Publisher: Multidisciplinary Digital Publishing Institute.

\bibitem{tu_effect_2019}
Zhiyong Tu and Changyong Xue.
\newblock Effect of bifurcation on the interaction between {Bitcoin} and
  {Litecoin}.
\newblock {\em Finance Research Letters}, 31, 2019.
\newblock Publisher: Elsevier.

\bibitem{garcia2020transfer}
Andr{\'e}s Garc{\'\i}a-Medina and Graciela Gonz{\'a}lez~Far{\'\i}as.
\newblock Transfer entropy as a variable selection methodology of
  cryptocurrencies in the framework of a high dimensional predictive model.
\newblock {\em PloS one}, 15(1):e0227269, 2020.

\bibitem{schreiber2000measuring}
Thomas Schreiber.
\newblock Measuring information transfer.
\newblock {\em Physical review letters}, 85(2):461, 2000.

\bibitem{barnett2009granger}
Lionel Barnett, Adam~B Barrett, and Anil~K Seth.
\newblock Granger causality and transfer entropy are equivalent for gaussian
  variables.
\newblock {\em Physical review letters}, 103(23):238701, 2009.

\bibitem{kraken}
Kraken downloadable historical market data (time and sales).
\newblock
  \url{https://support.kraken.com/hc/en-us/articles/360047543791-Downloadable-historical-market-data-time-and-sales},
  2022.

\bibitem{britannica}
The Editors of~Encyclopaedia Britannica.
\newblock fiat money.
\newblock \url{www.britannica.com/topic/fiat-money}, 2022 (accessed 28 March,
  2022).

\bibitem{hamilton2020time}
James~Douglas Hamilton.
\newblock {\em Time series analysis}.
\newblock Princeton university press, 2020.

\bibitem{geweke_measurement_1982}
John Geweke.
\newblock Measurement of linear dependence and feedback between multiple time
  series.
\newblock {\em Journal of the American statistical association},
  77(378):304--313, 1982.
\newblock Publisher: Taylor \& Francis.

\bibitem{barnett_mvgc_2014}
Lionel Barnett and Anil~K. Seth.
\newblock The {MVGC} multivariate {Granger} causality toolbox: a new approach
  to {Granger}-causal inference.
\newblock {\em Journal of neuroscience methods}, 223:50--68, 2014.
\newblock Publisher: Elsevier.

\bibitem{watanabe_information_1960}
Satosi Watanabe.
\newblock Information theoretical analysis of multivariate correlation.
\newblock {\em IBM Journal of research and development}, 4(1):66--82, 1960.
\newblock Publisher: IBM.

\bibitem{sun_linear_1975}
T.~H. Sun.
\newblock Linear dependence structure of the entropy space.
\newblock {\em Inf Control}, 29(4):337--68, 1975.
\newblock Publisher: Elsevier.

\bibitem{scagliarini_quantifying_2022}
Tomas Scagliarini, Daniele Marinazzo, Yike Guo, Sebastiano Stramaglia, and
  Fernando~E. Rosas.
\newblock Quantifying high-order interdependencies on individual patterns via
  the local {O}-information: theory and applications to music analysis.
\newblock {\em Physical Review Research}, 4(1):013184, 2022.
\newblock Publisher: APS.

\bibitem{stramaglia_quantifying_2021}
Sebastiano Stramaglia, Tomas Scagliarini, Bryan~C. Daniels, and Daniele
  Marinazzo.
\newblock Quantifying dynamical high-order interdependencies from the
  o-information: an application to neural spiking dynamics.
\newblock {\em Frontiers in Physiology}, 11:1784, 2021.
\newblock Publisher: Frontiers.

\bibitem{schwarz_estimating_1978}
Gideon Schwarz.
\newblock Estimating the dimension of a model.
\newblock {\em The annals of statistics}, pages 461--464, 1978.
\newblock Publisher: JSTOR.

\end{thebibliography}

\end{document}